%% file: main.tex
\documentclass[letterpaper,10pt]{article}   
\usepackage{authblk}

\usepackage[letterpaper,margin=1in]{geometry} 
\usepackage{setspace}
\usepackage[T1]{fontenc}
\usepackage{lmodern}     

\usepackage{graphicx}
\usepackage[table,xcdraw]{xcolor}

\usepackage[hidelinks]{hyperref}

\usepackage{array}
\usepackage{booktabs}

\usepackage{tabularx}
\usepackage{multirow}
\usepackage{makecell}

\usepackage{orcidlink}
\usepackage{svg}

\usepackage{listings,lipsum}
\usepackage{color}

\usepackage{todonotes}
\usepackage{xspace}
\usepackage{xifthen}
\usepackage{glossaries}

\usepackage{pifont} 
\usepackage{wasysym}
\usepackage{enumitem}
\usepackage{amssymb}   
\usepackage{amsmath}   
\usepackage{fontawesome}


\usepackage{soul}
\usepackage{xcolor}

\definecolor{codegreen}{rgb}{0,0.6,0}
\definecolor{codegray}{rgb}{0.5,0.5,0.5}
\definecolor{codepurple}{rgb}{0.58,0,0.82}
\definecolor{backcolour}{rgb}{0.95,0.95,0.92}

\lstdefinestyle{cstyle}{
    language=C,
    backgroundcolor=\color{backcolour},
    commentstyle=\color{codegreen},
    keywordstyle=\color{codepurple},
    numberstyle=\tiny\color{codegray},
    stringstyle=\color{codegreen},
    basicstyle=\ttfamily\small,
    breakatwhitespace=false,
    breaklines=true,
    captionpos=b,
    keepspaces=true,
    numbers=left,
    numbersep=5pt,
    showspaces=false,
    showstringspaces=false,
    showtabs=false,
    tabsize=2
}
\input{glossary}

\usepackage{xspace}

\newcommand{\binsec}{\textsc{Binsec/Rel2}\xspace}
\newcommand{\ctgrind}{\texttt{ctgrind}\xspace}
\newcommand{\dudect}{\texttt{dudect}\xspace} 
\newcommand{\flowtracker}{{FlowTracker}\xspace}
\newcommand{\rtlf}{\textsc{RTLF}\xspace}
\newcommand{\timecop}{\textsc{TIMECOP}\xspace}

\title{Systematic Timing Leakage Analysis of NIST PQDSS Candidates: \\ Tooling and Lessons Learned}

\author[1]{Olivier Adjonyo\orcidlink{0000-0002-3863-3714}} 
\author[2]{Sébastien Bardin\orcidlink{0000-0002-6509-3506}}
\author[1]{Emanuele Bellini\orcidlink{0000-0002-2349-0247}} 
\author[1]{Gilbert Ndollane Dione\orcidlink{0000-0001-8583-0668}}
\author[2]{Mahmudul {Faisal Al Ameen}\orcidlink{0000-0002-2388-4908}}
\author[1]{Robert Merget\orcidlink{0000-0003-2618-1156}}
\author[2]{Frédéric Recoules\orcidlink{0009-0001-9736-0656}}
\author[2]{Yanis Sellami\orcidlink{0009-0006-8833-3863}}

\affil[1]{Technology Innovation Institute, Abu Dhabi, UAE}
\affil[2]{Université Paris-Saclay, CEA, List, F-91120, Palaiseau, France}

\date{\ }

\begin{document}

\maketitle

\vspace{-0.5cm}

\begin{abstract}
The PQDSS standardization process requires cryptographic primitives to be free from vulnerabilities, including timing and cache side-channels. Resistance to timing leakage is therefore an essential property, and achieving this typically relies on software implementations that follow constant-time principles.  
Moreover, ensuring that all implementations are constant-time is crucial for fair performance comparisons, as secure implementations often incur additional overhead.  
Such analysis also helps identify scheme proposals that are inherently difficult to implement in constant time.  
Because constant-time properties can be broken during compilation, it is often necessary to analyze the compiled binary directly.  
Since manual binary analysis is extremely challenging, automated analysis becomes highly important.  
Although several tools exist to assist with such analysis, they often have usability limitations and are difficult to set up correctly.  
To support the developers besides the NIST committee in verifying candidates, we developed a toolchain that automates configuration, execution, and result analysis for several widely used constant-time analysis tools.  
We selected \timecop and \binsec to verify constant-time policy compliance at the binary level, and \dudect and \rtlf to detect side-channel vulnerabilities through statistical analysis of execution time behavior.  
We demonstrate its effectiveness and practicability 
by evaluating the NIST PQDSS round 1 and round 2 implementations.    
We reported 26 issues in total to the respective developers, and 5 of them have already been fixed. We also discuss our different findings, as well as the benefits of shortcomings of the different tools.  

\begin{paragraph}{Keywords}
    Cryptographic library, Side-Channel, Time leakage, Automated constant-time analysis tool, Post-Quantum, Digital Signature
\end{paragraph}
\end{abstract}

\section{Introduction}
\label{sec:intro}
The National Institute of Standards and Technology (NIST) has led efforts to transition the cryptographic community toward quantum-resistant algorithms in response to the growing threat posed by quantum computing. 
Through a multi-round public standardization process initiated in 2016 \cite{nist_pqc}, NIST selected key encapsulation mechanisms (CRYSTALS-KYBER~\cite{FIPS-203}, HQC~\cite{nist-pqc-selected}) and digital signature algorithms (CRYSTALS-Dilithium~\cite{FIPS-204}, FALCON~\cite{nist-pqc-selected}, SPHINCS+~\cite{FIPS-205}), most of which are based on structured lattices. 
Due to limited alternatives in lattice-free digital signatures, NIST launched a subsequent call in 2022 for additional proposals~\cite{pqc-dig-sig}, leading to a second round evaluation of 14 candidates. All candidates must demonstrate resistance to side-channel attacks~\cite{NIST-IR-8528,nist-pqc-dig-sig-cfp-2022,nist-pqc-security-criteria}, particularly by following constant-time (CT) principles such as avoiding secret-dependent control flows and memory accesses. The importance of analyzing side-channel vulnerabilities in post-quantum cryptographic (PQC) schemes has been well-documented~\cite{timingntruencrypt2007,cryptoeprint:2016/300,cryptoeprint:2017/505}, with notable attacks such as KyberSlash~\cite{kyberslash} revealing timing leaks in widely used implementations. To support developers, a variety of constant-time analysis tools have been developed, including 
tools that aim to detect general timing differences/attacks, such as \emph{statistical} tools (e.g. \dudect~\cite{dudect16}, mona~\cite{mona12}, tlsfuzzer~\cite{tlsfuzzer23} or \rtlf~\cite{rtlf}), or
tools that aim to detect violations of the constant-time policy, such as 
\emph{static formal} tools (e.g. BLAZER~\cite{blazer}, CacheAudit~\cite{cacheaudit13}, \flowtracker~\cite{flowtracker} or CT-LLVM~\cite{ctllvm}), 
\emph{dynamic} tools (e.g. \ctgrind~\cite{ctgrind10}, MicroWalk~\cite{microwalk18}, DATA~\cite{data18} or \timecop~\cite{timecop}), and 
\emph{symbolic} tools (e.g. \binsec~\cite{binsecrel19} or CacheD~\cite{cached17}), 
each targeting different aspects of side-channel detection in cryptographic code.

All the aforementioned tools' application to post-quantum cryptographic (PQC), or, more generally, to cryptographic implementations remains limited. 
A study by Jancar et al.~\cite{notthathardtomitigate21} revealed that most cryptographic software developers do not fully utilize these tools, often citing poor usability as a key barrier. Contributing factors include inconsistencies in tool configurations, difficulty creating appropriate configuration files, and timing-based leakages that are not even captured. Moreover, specific techniques struggle to scale effectively to cryptographic use cases, and the resulting analysis reports are often too complex to interpret. Geimer et al.~\cite{GeimerVRDBM23} evaluated the efficacy of various tools on classical cryptographic primitives and concluded with specific technical recommendations, calling for closer collaboration between tool developers and maintainers of cryptographic libraries.

Our goal is to enable the systematic use of binary-level side-channel analysis tools for evaluating implementations of post-quantum cryptographic (PQC) algorithms, focusing on improving their ability to report relevant and actionable findings to developers. 

To address the current limitations, we propose leveraging the standardization efforts led by NIST to design an automated toolchain tailored for PQC implementations. This toolchain will integrate existing binary-level timing side-channel detection tools and extend their applicability to post-quantum digital signature algorithms (DSA) implementations. By doing so, we aim to identify and report potential timing side-channel vulnerabilities more effectively, ultimately supporting developers in building more secure cryptographic software.

As the standardization process moves forward, we expect the focus to shift more towards side-channel resistant implementations, and we hope that our efforts in selecting and integrating side-channel analysis tools in an easy-to-use toolchain will allow authors to better integrate timing leakage testing into their development process.

\paragraph{Contributions} Our main contributions are the following: 

\begin{itemize}

\item To assist both candidates and NIST, we propose a toolchain of popular open-source timing-leakage analysis tools covering three types of time analysis approaches across four tools: 
\timecop~\cite{timecop}\footnote{\url{https://www.post-apocalyptic-crypto.org/timecop}},
\binsec~\cite{binsecrel19}\footnote{\url{https://github.com/binsec/binsec}}, \dudect~\cite{dudect16}\footnote{\url{https://github.com/oreparaz/dudect}} and \rtlf\cite{rtlf}\footnote{\url{https://github.com/tls-attacker/RTLF}}.
The toolchain allows for the automation of the process of detecting binaries with timing side-channel vulnerabilities and code where constant-time policy is violated, in particular by automatically generating test harness and configuration files required by each tool.
We provide a public Docker image of the toolchain.

\item In order to demonstrate the interest and effectiveness of our toolchain, we integrated Round 1 and Round 2 candidates of the NIST PQDSS standardization process in the toolchain and performed a systematic evaluation on all of their instances. Our method allowed us to analyze 302 instances of 15 primitives. We then manually reviewed a subset of the reported issues (\autoref{sec:findings}).
After a manual pruning of the tool alerts, we reported the 26  most critical issues to the respective authors, among which \textit{5 have been already acknowledged as critical vulnerabilities and fixed} by their developers, namely the alerts from SNOVA and Preon from round 1 and LESS (2 alerts) and Mirath from round 2.  

\item We further evaluated on selected candidates how the different tools compare and complement each other—for example, using a precise but slower tool like \binsec to confirm results from faster tools like \timecop, which produces significantly more false positives, or to pinpoint issues flagged by statistical tools like \dudect or \rtlf that only analyze overall execution time. In particular, we performed a detailed comparison between  \binsec and \timecop, showing that the tools produce a set of overlapping findings and some distinct findings as well. While \timecop runs faster, has better code coverage, and flags more issues, \binsec achieves a much higher precision with respect to its constant-time policy. 
In particular, on a small subset of benchmarks for which we manually checked, \binsec always raises alerts that are actual constant-time violations, while 67\% of 
\timecop alerts turned out not to be actual constant-time violations. 

\end{itemize}

\noindent \textit{
All the material to reproduce our results, including a Docker image, NIST candidates, examples, and documentation, is publicly available at: }
\begin{center}
\texttt{\url{https://github.com/Crypto-TII/pqdss-ct-toolchain}}     
\end{center}


\section{Background}

\label{sec:scct}

Timing side-channel analysis is a pivotal area of research within the domain of computer security, focusing on exploiting subtle variations in the execution time of cryptographic algorithms to extract sensitive information. The origins of this field can be traced back to Paul Kocher\cite{timingRsaDhKocher96}, who pioneered the concept of side-channel attacks. Regarding timing side-channel vulnerabilities, cryptographic operations should not have variable execution time that correlates to secret values. To fight these issues, the community established a set of programming best practices known as \textit{Constant-time Programming}\footnote{A misnomer as this only requires branches and memory accesses to be independent of secret values.}, which provides rules a programmer should follow to help avoid a large collection of potential timing attack vulnerabilities.
The usual constant-time policy states that no branching condition and no memory access should depend on secret values, the rationale being that it prevents distinct path timing differences and cache-related timing vulnerabilities
\cite{DBLP:conf/ccs/ShivakumarBGLP22,DBLP:conf/uss/DanielBNBRP23}. 
In practice, however, more leakage sources exist that affect the real execution time of code on actual hardware. For example, on an ARM Cortex M3 processor, the \texttt{UMUL} instruction has variable length execution time depending on the value of the operands~\cite{groot15}. Similarly, on many CPUs, division instructions also have variable-length execution times, which were recently exploited in the Kyberslash attack~\cite{kyberslash}. Another example of non-constant-time behavior is the impact of CPU frequency scaling on the execution time as discovered by Wang et al.~\cite{hertzbleed22}. Concretely, Wang et al. showed that the hamming weight of used operands affects the temperature of the CPU, which can cause the CPU to slow down due to dynamic CPU frequency scaling to avoid overheating. This slowdown can be measured by an attacker, who then leaks the Hamming weight of the involved secrets.

Whether these leakage sources should be considered when writing code depends on the targeted security level and hardware.

\subsection{Timing Side-Channel Detection Tools}
\label{sec:cttools}
To help developers detect potential timing differences in execution time coming from secret values, the community has developed a large collection of tools.
Jancar et al.~\cite{notthathardtomitigate21} group these tools into four categories: formal, dynamic, symbolic, and statistical.

\subsubsection{Formal}

Formal tools
(e.g. BLAZER~\cite{blazer}, CacheAudit~\cite{cacheaudit13}, \flowtracker~\cite{flowtracker} or CT-LLVM~\cite{ctllvm})
try to reason for or against the leakage by looking at the source code, LLVM IR code, or the produced binary -- usually, they can, in principle, prove the absence of leakage (possibly with some help from the user) but cannot generate witnesses of violations and may suffer from a high false-positive rate. For a false-positive-free evaluation, formal tools need to know which values are considered secret and public. The respective values, therefore, have to be annotated by the user of the tool.

\subsubsection{Dynamic}

With dynamic techniques (e.g., \ctgrind~\cite{ctgrind10}, MicroWalk~\cite{microwalk18}, DATA~\cite{data18} or \timecop~\cite{timecop}), the source code is unavailable to the tool. Instead, dynamic tools execute the code, track how secret values are handled at runtime, and try to pinpoint timing leaks. A drawback of this approach is that the code that causes a leak has to be executed, which is not always trivially achievable if the code is part of a specific corner case that is only triggered based on a specific public value. As with formal tools, dynamic tools have a specific leakage model in mind and require that the user annotates secret values.

\subsubsection{Symbolic}

Symbolic tools (e.g., \binsec~\cite{binsecrel19} or CacheD~\cite{cached17}) are a middle ground between formal and dynamic techniques. They precisely execute the code symbolically to argue about the reachability of the code and violations in order to reduce the number of false results, often with the ability to generate witnesses of violation. As a downside, they can only explore a finite number of paths, yielding bounded formal proofs at best. As with formal and dynamic methods, symbolic techniques also require that secret values are annotated and that a leakage model is defined (either by the user or the tool itself). 

\subsubsection{Statistical}

In contrast to the previously mentioned tools, statistical tools (e.g. \dudect~\cite{dudect16}, mona~\cite{mona12}, tlsfuzzer~\cite{tlsfuzzer23} or \rtlf~\cite{rtlf}) do not require the user (or tool) to carefully model the potential sources of leakage but instead, simply measure the execution time of the code and then try to make a statistical argument on whether the code is executing in secret independent time. 
This argument is statistical and will never be sound or complete. 
Another disadvantage of (current) statistical tools is that they cannot perform fault localization, leaving it up to the developer to find the actual leakage in the code. 
Since statistical tools measure execution time, they can also detect leakages that are due to the underlying hardware architecture and not to the code itself.
Statistical tools should be used in combination with other tools as a "last line of defense" to really check that when executed on real hardware, indeed, no leakage is detected.

\section{Related Work}
\label{sec:relatedwork}

A similar effort to the one reported in this manuscript is the report written by Hansen et al. \cite{LW-toolchain}, collecting a set of side-channel analysis tools (\flowtracker, \dudect, and \ctgrind) to evaluate all 32~candidates in the second round of the NIST lightweight cryptography standardization process.
Our work is similar but targets the PQDSS standardization process instead and makes use of more recent side-channel analysis tools.

A similar large-scale study was conducted with \timecop~\cite{timecop}, which used a \ctgrind style approach to analyze implementations from the SUPERCOP toolkit~\cite{supercop}.
By contrast, we perform a cross-evaluation between several tools in this work.

\timecop was recently extended with the ability to detect side-channels from variable time instructions~\cite{cryptoeprint:2024/1049}.
At the time of writing, only the CROSS team has submitted its implementation for evaluation in SUPERCOP \cite{supercop-sign-benchmarks}, in spite of the February 2025 public call to submit raised in the Google pqc-forum group \cite{nist-pqc-forum-2024}.

\section{PQDSS Time Leakage Toolchain}
\label{sec:toolchain}
All NIST PQDSS candidates have to implement the same API for signing, key generation and signature verification.
To help with systematic testing of submissions to the NIST PQDSS standardization process for timing leakages, we created a toolchain that uses prominent tools for timing side-channel analysis that specifically targets this API. 

\subsection{Toolchain overview}

We evaluate the candidates on the same API specified by NIST. We only analyze the \texttt{crypto\_sign()} function from the main \texttt{api.h} file. For all the tests, we consider the private key as the only secret data. The rest of the inputs are considered public. 

Our toolchain does not provide the required configuration files statically but generates them on demand. To perform tests with a given tool on a given candidate, one has to invoke a Python script, which will generate the necessary configuration and run the tool. 
The toolchain performs the analysis in four phases:

\begin{enumerate}
    \item Generates the required configuration file and test harness (.c, .ini) according to the chosen algorithm and tool.

    \item Compiles the binary of the generated test harness, taking into account the tools' compilation/link flags.
    \item Run the tool on that binary.

\end{enumerate}

\begin{figure}
    \centering
    \includegraphics[width=\linewidth]{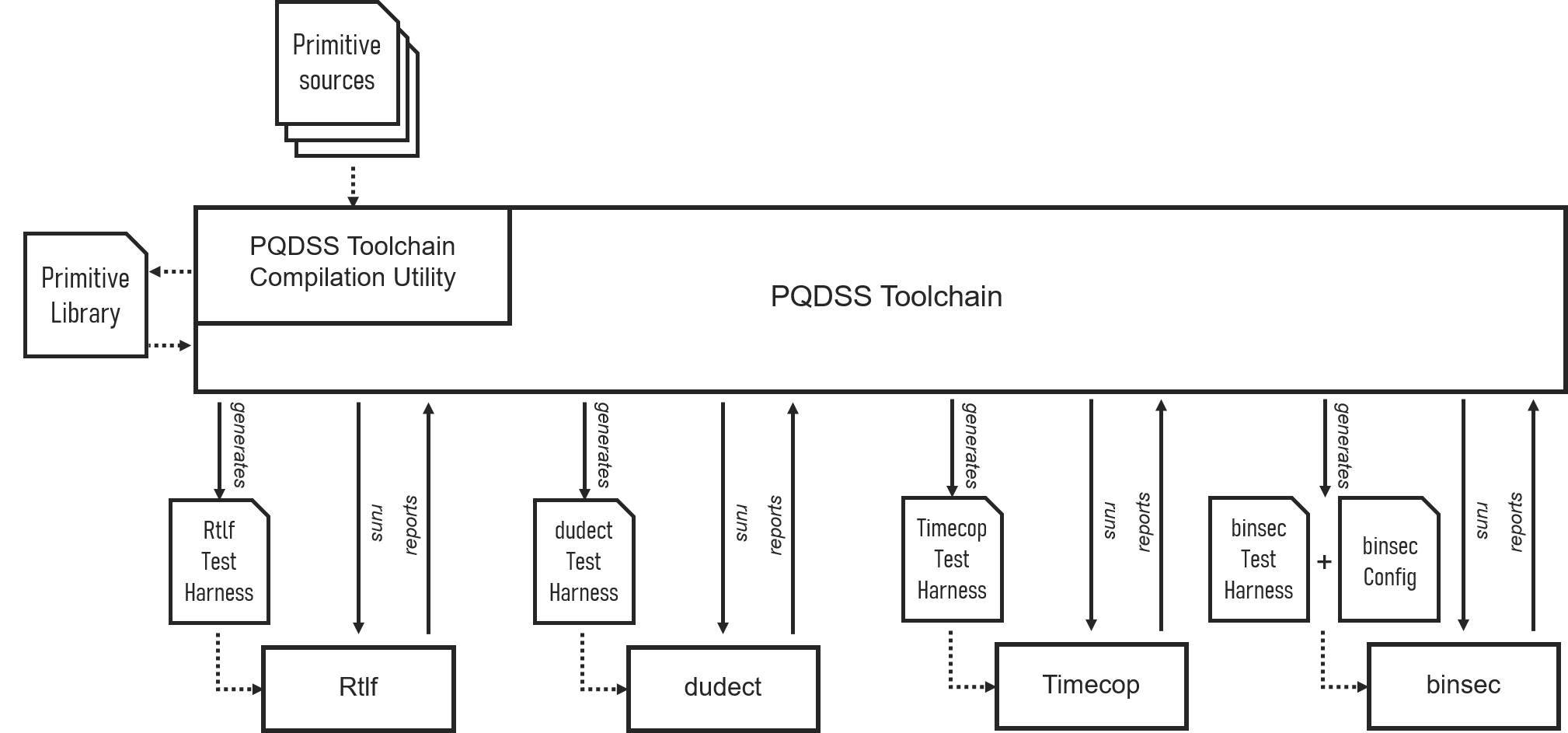}
    \caption{Graphical Representation of the Toolchain}
    \label{fig:toolchain}
\end{figure}

A graphical representation of the toolchain is available in \autoref{fig:toolchain}.

\subsection{Tools and Configurations}

In this study, we are only interested in tools that evaluate full binary implementations.
We will differentiate between two groups: the tools that aim to help developers prove that their code follows a constant-time (CT) policy, and the tools that perform a statistical test on the execution time of the algorithm on actual hardware to check whether it correlates with secret values.
In all cases, tools were selected after considering their popularity, successful applications to detect timing leakages, and the status of development.

\subsubsection{\timecop}
\label{sec:timecop}

We chose the dynamic tool \timecop~\cite{timecop} as our first constant-time policy tool. 
\timecop is essentially a repackaging of Langley's approach~\cite{Langley2010checkvalgrind}, applied to cryptographic algorithms in the SUPERCOP benchmarking suite, and publishing its results.
Our \timecop uses Valgrind 3.23.0\footnote{\url{https://valgrind.org/}} to detect potential timing dependence on secret data by marking secret data as uninitialized memory. 
\timecop is neither sound  (it follows the execution trace) nor complete (it taints memory blocks) for the constant-time policy it targets.

We classify the private key as secret, run the signing function \texttt{crypto\_sign()}, and then declassify it again.
Note that we do not classify (\texttt{poison()}) nor declassify (\texttt{unpoison()}) values within the analyzed function, which can lead to false positives in the presence of masking, for instance.
We run the tests for a single iteration and with no timeout for the execution. 

\subsubsection{\binsec}
\label{sec:binsec}

As our second constant-time policy tool, we chose the symbolic analysis tool \binsec~\cite{binsecrel19,DanielBR23}. The tool works at the binary level and performs a \textit{relational} symbolic execution in order to detect constant-time violations. A witness of violation is generated in case such a violation is found; otherwise, the program is proven correct up to a given instruction count bound (which for cryptographic code, often translates into a given key size). We take the latest version of the tool at the time of writing, described in \cite{GeimerVRDBM23}. Our toolchain uses \binsec 0.10.0.

Initializing the symbolic engine is a notoriously tedious task: the initial state must contain consistent values for pointers, code addresses, dynamically linked libraries, etc.
We initialize the memory at the call to \texttt{crypto\_sign} with a core dump obtained by executing the test harness up to this call in GDB.
We then overwrite the content of the public and secret buffer variables with a symbolic variable of the corresponding taint.
We opted for a bug-finding targeted setup by removing the nondeterminism of the public data (concrete values) and by limiting the private taint to a single byte of the key.
Additionally, note that allocation functions and AES instructions have been stubbed.
This breaks the completeness of the analysis, which means that the absence of reported violations will not imply the absence of constant-time violations.

For all instances, we run \binsec with a 30-minute timeout and also request alerts on secret-dependent operands of multiplication and division instructions, as recent work~\cite{kyberslash} has demonstrated their importance.

\subsubsection{\dudect \& \rtlf}

For the statistical tools, we chose \dudect~\cite{dudect16} and \rtlf\cite{rtlf}.
\dudect measures the number of CPU cycles executed during program execution and then uses a series of Welch’s t-tests \cite{Sachs.1984} on the gathered data. The measurement setup needs to be unbiased; otherwise, measurement artifacts will be reported as time leakages. Our toolchain uses \dudect\texttt{:a18fdee2}
and \rtlf\texttt{:2497d66}. 
The measurements from \dudect can be exported to be used in \rtlf. 
\rtlf implements statistical hypothesis testing using a combination of quantile estimators with an empirical bootstrap, with a configurable false positive rate.
Compared to \dudect, this allows the user for a more robust analysis, also in the presence of noise.

\label{sec:dudect}
To test if an implementation contains a statistically measurable side-channel with \dudect and \rtlf, we need to test inputs that are drawn from two different classes. The null hypothesis is that the runtime of inputs from either class will not result in a difference in the runtime of the algorithm.
We chose to measure random private keys.
For each scheme, we then expanded the measured private key into its mathematical representation and then sorted the keys into two different input classes. We then created classes for the inputs based on a specific bit in the expanded key. We once selected a bit at the beginning of the key, one in the middle of the key, and one bit at the end of the key. The specific bits were chosen to try to create input classes that are roughly balanced, such that the algorithm will have roughly the same number of measurements for both classes. Balancing out the distribution was not always possible with this generic approach due to their mathematical structure.  

While \dudect does not allow for setting a false positive rate,
for \rtlf, we used a false positive rate of 9\%, corresponding to a significance level $\alpha=0.09$.

To minimize external noise in timing measurements, we avoid Docker and instead evaluate the system directly on bare-metal hardware.
To further reduce interference, we reserve specific CPU cores via GRUB configuration and schedule processes individually using \texttt{taskset}.
The measurements are obtained in one shot for both \dudect and \rtlf. Then the statistics are computed in a second step.
We set a default number of measurements of 210k for \dudect and \rtlf, to ensure that for balanced bits, we get roughly 100k measurements for each class.

\subsection{Limitations}

As of now, we treat the respective algorithms as a black box and have not yet adjusted our tools to the internals of the respective algorithms. For example, after the initial classification of secrets, we do not declassify secrets if they are masked adequately within an algorithm implementation, which can cause the used tools to flag CT violations that are not relevant to the user.

Another common source of irrelevant CT violations is found in algorithms using rejection sampling. In this case, the algorithm samples values from a secret and then discards them if the sampled values do not have a specific structure. This inherently creates a variable execution time, as a different number of sampling operations are performed based on a secret. This variable execution time is usually not exploitable, as the secret is typically not used directly. For example, implementations may hash the secret together with a round counter and then sample from the hash. which at most leaks information about the output hash, which is not used anymore if the resulting sampled data does not have a desired property.

Another limitation arises from how the secret key is modeled. In many cryptographic schemes, the secret key structure includes both secret and public information, such as public seeds, public key hashes, or other derived constants. However, often constant-time tools treat the entire secret key buffer as uniformly sensitive. This coarse-grained sensitivity model can lead to false positives when implementations access the public portion of the secret key in a data-dependent way. Ideally, constant-time checkers should support more fine-grained annotations or semantic awareness to distinguish between truly sensitive and public fields within a compound key structure.

\section{Evaluation}
\label{sec:findings}

We evaluate our toolchain in the context of the NIST PQDSS candidates analysis.
The primitives chosen for integration are listed in \autoref{tab:results}.
In all cases, we performed the evaluation on the Optimized Implementation submitted to the NIST PQDSS selection process.
In particular, consider the following three research questions:
\begin{enumerate}
    \item[\textbf{RQ1}] Can we leverage the toolchain to detect vulnerabilities?
    \item[\textbf{RQ2}] How do symbolic and statistical tools synergize?
    \item[\textbf{RQ3}] What is the quality of the symbolic tools' alerts?
\end{enumerate}

\subsection{\textbf{RQ1}: Toolchain findings}
\label{sec:rq1}

In this section, we highlight the findings obtained from the results reported by the aforementioned tools on the post-quantum digital signature schemes.
This includes a first analysis of the alerts raised by the tools, a systematic and responsible disclosure of the discovered potential constant-time violations, and a tracking of the results of this disclosure.

\subsubsection{Tool Results: Alerts Analysis}
\label{sec:alerts}

\input{comparison_table1}

In \autoref{tab:results}, we mark the outcome of our analysis, for each primitive instance and for each tool, as follows:

\begin{itemize}
    \item[\nointegration] -- when we could not compile the primitive instance
    \item[\notevaluated] -- when no experiment was run

    \item[\noalert] -- when the tool did not report any alerts (including timeout with no alerts)
    \item[\notrelevant] -- when all alerts were found not to be constant-time issues 
    \item[\rawalert] -- when at least one alert was not investigated
    \item[\reported] -- when at least one alert was considered a real constant-time issue and reported to the developers
    \item[\truepositive] -- when at least one alert was considered a real constant-time issue and acknowledged by the developers
\end{itemize}

We detail the conclusions of our findings below, organized per tool.

\paragraph{\timecop}
\label{sec:find_timecop}

\timecop reported constant-time violations in all the implementations.
However, some of those issues are related to the public part of the secret key.
Indeed, as the whole secret key is poisoned as a single block, the whole corresponding memory region is considered to contain secret data.
Thus, any secret-dependent conditional branch and memory access makes the tool flag a constant-time violation, even if it comes from the public data of the secret key.
We consider these issues as false positives. Besides, we identified issues where the branching never varies across executions due to constraints on the secret data, which collectively give a high rate of false positives.
\timecop also reported eight issues in the implementations of rejection sampling. 

\paragraph{\binsec}
\binsec raised an alert on at least one instance of all primitives except for MQOM, SQIsign, SDitH, and UOV.
In all instances, the exploration was not completed, and the analysis was either cut by a timeout trigger or as a result of memory exhaustion.
In all timeout triggers, we identified that \binsec was stuck waiting for an SMT query answer, which is likely a consequence of the computational complexity of inverting the algorithms under study.
In the cases where analysis was stopped due to memory exhaustion, \binsec struggled because its formula representation and simplification strategies were not well-suited to certain code patterns. This led to scalability issues, especially in computation-heavy code such as the matrix multiplication routines in SNOVA.
This also explains why, for SNOVA and CROSS, a constant-time violation present in their shared source code—likely affecting all parameter instances—is not detected by \binsec in some cases. 
The increased computational complexity introduced by higher parameter values leads to early termination of the analysis, preventing the violation from being reported.
This explains its inconsistent detection across instances compared to simpler cases. 

Among all the instances for which \binsec raised alerts, three of them were acknowledged by the developers as exploitable and in need of a fix. 
The others are valid constant-time violations, but they arose on a part of the implementation that was not expected to be constant-time: either on part of the code for which the hardening has not been done yet or because the developers believe that randomness of other components of the conditional/address computation will prevent its exploitation in practice (see also \autoref{sec:noexploit}). 

\paragraph{\dudect and \rtlf}
\label{sec:find_dudect}
\label{sec:find_rtlf}

As both \dudect and \rtlf only report alerts on the whole program, not for a specific location, we cannot manually investigate the relevancy of the alert.
This means that the most serious alert status for both tools is \rawalert.
This permits the distinction between instances that were found to have a statistically significant timing dependency on secret data from those that were not.

\subsubsection{Responsible Disclosure}
\label{sec:disclosure}

In order to validate the potential sources of vulnerabilities (alerts) that we detected, we performed a responsible disclosure of our findings to the teams developing the corresponding digital signature primitives.

We disclosed manually confirmed constant-time violations to the developers of SNOVA (rounds 1 and 2), Preon (round 1), LESS, MAYO, Mirath, CROSS, HAWK, QR-UOV, RYDE, and PERK (round 2), receiving responses from the developers of MAYO, Mirath, CROSS, SNOVA, QR-UOV, RYDE, and PERK. 
The disclosure was done with a report detailing the evaluation setup and our understanding of the potential vulnerability.
The detailed description of all the acknowledged vulnerabilities is available in \autoref{sec:vulns}.
We use their answers as the ground truth for exploitability/non-exploitability results in \autoref{tab:results}.

\input{comparison_table2_nist}

\autoref{tab:report} details the disclosure status, listing for each tool the total number of alerts raised across all instances of a primitive at the binary level, the number of distinct source locations that these binary alerts correspond to, and the number of reported alerts—manually verified from a subset of these locations with false positives removed—forwarded to the schemes’ developers.
\autoref{tab:report} additionally shows the developer acknowledgment status of these alerts.

\subsubsection{From CT Violation to Vulnerability}
\label{sec:vulns}

Across the tested candidates, we could identify multiple vulnerabilities in both their round 1 and round 2 versions. We did not further analyze whether the leakage is serious enough to actually recover key material.
Instead, we considered a weaker notion of \emph{acknowledgment} by the primitives' developers.
We report in this section the vulnerabilities that were confirmed to be true constant-time violations and acknowledged as \emph{requiring to be fixed} by the developers.

\paragraph{SNOVA} For the generation of the $T^{12}$ value, the SNOVA implementation used a secret-based memory lookup, which could leak parts of the scaling factor $k$ to an attacker with access to cache timings.

The issue reported by \binsec concerns the generation of \texttt{T12} in the function \texttt{gen\_seeds\_and\_T12} from the secret \texttt{seed}. During the computation of \texttt{T12}, memory is accessed with the indices that are derived from \texttt{seed} and thus is secret, which can enable cache-based side-channel attacks in this implementation.
For example, in the instance \texttt{SNOVA-24-5-16-4-esk}, the function \texttt{gen\_{a\_FqS}} is called in snova.c at line 215 where \texttt{pt\_array} is secret, which in turn calls the function \texttt{gf16m\_scale} in snova.c at line 187. In this function, the parameter \texttt{c} is thus secret, and when \texttt{c[i]} is passed as \texttt{k} to the function \texttt{gf16\_scale}, it is used as an index in the following expression:

\begin{lstlisting}[style=cstyle, label=lst:snova, numbers=none]
c[(i << 2) ^ j] = mt4b[(a[(i << 2) ^ j] << 4) ^ k]
\end{lstlisting}

In a co-located attacker model, the index used in any memory access has to be considered public, meaning that the above memory access can be leaked to the attacker, which depends on \texttt{k}.

\paragraph{Preon} The AES implementation that is used within the Preon round 1 code contained an insecure S-Box implementation which relied on a secret-based table lookup. This allows an attacker to deduce the inputs to the S-Box via cache timing behavior.

The issue reported by \binsec concerns the AES implementation that is bundled with the scheme. The AES implementation being used implements the S-box of AES with a table lookup, which introduces a cache side-channel vulnerability in the code.
Here we briefly show the vulnerable code in the optimized implementation of \texttt{Preon128A}, which is expected to be free from constant-time violations. From the following code in aes.c at line 184, we observe that the index \texttt{num} is used to access a cell of \texttt{sbox}. 

\begin{lstlisting}[style=cstyle, caption={Vulnerable S-box lookup in AES}, label=lst:taint_aes_sbox, numbers=none]
tempa[0] = getSBoxValue(tempa[0]);
static uint8_t getSBoxValue(uint8_t num) {
    return sbox[num];
}
\end{lstlisting}

The parameter \texttt{num} corresponds to \texttt{tempa[0]}, which is directly dependent on the secret key. Thus, the above memory access may leak the secret data to an adversary via cache-based side channel attacks.

\paragraph{LESS} 

    During the commitment step,  monomial matrices  $\Tilde{\mu}$ are generated and are used as commitments. Those matrices are supposed to be secret before the revealing/response phase, at least before the prover sends them, as a response to the verifier, depending on the challenge. For the generation of $\Tilde{\mu}$, the LESS implementation used secret-based memory access.


\timecop reported the leakage in the function \texttt{yt\_shuffle\_state} in monomial.c at line 90  shown in the following listing.

The secret positions \texttt{x} of the elements of the permutation are leaked through a memory index access by \texttt{permutation[i] = permutation[x];}.
The same issue applies at line 91.
        
\begin{lstlisting}[style=cstyle, caption={function: \texttt{yt\_shuffle\_state}}, label=lst:taint_rnd, firstnumber=65]
void yt_shuffle_state(SHAKE_STATE_STRUCT *shake_monomial_state, POSITION_T permutation[N]) {
   uint64_t rand_u64;
   POSITION_T tmp;
   POSITION_T x;
   int c;

   csprng_randombytes((unsigned char *) &rand_u64,
                             sizeof(rand_u64),
                             shake_monomial_state);
   c = 0;

   for (int i = 0; i < N; i++) {
      do {
         if (c == (64/POS_BITS)-1) {
            csprng_randombytes((unsigned char *) &rand_u64,
                                sizeof(rand_u64),
                                shake_monomial_state);
            c = 0;
         }
         x = rand_u64 & (POS_MASK);
         rand_u64 = rand_u64 >> POS_BITS;
         c = c + 1;
      } while (x >= N);

      tmp = permutation[i];
      permutation[i] = permutation[x];
      permutation[x] = tmp;
   } 
}
\end{lstlisting}

Note that \texttt{shake\_monomial\_state} is tainted as secret by \texttt{sk\_shake\_state} via the ephemeral monomial seeds in the function \texttt{LESS\_sign} in LESS.c. Hence \texttt{rand\_u64} becomes secret dependent by the function \texttt{csprng\_randombytes}, which has  \texttt{shake\_monomial\_state} as an input parameter. 
 \texttt{x} is then secret as being derived from \texttt{rand\_u64}.

\paragraph{Mirath}

The alert reported by \binsec and also confirmed by \timecop concerns a memory access leak in the implementation of the matrix product, which is used to compute the product between the \texttt{S} and \texttt{C} matrices (built from secret values).
It accesses a precomputed result table at an index computed from the matrices' secret elements.
This may generate a cache timing attack.
The issue was confirmed by the developers and fixed in a subsequent code update.


From the key generation function \texttt{mirath\_keygen}, we observe that the first sixteen bytes of the parameter \texttt{sk} are treated as the secret key. Accordingly, we assume that the same bytes of \texttt{sk} are secret in the signing function \texttt{mirath\_sign} as well.

\begin{lstlisting}[style=cstyle, caption={Construction of the secret matrices \texttt{S} and \texttt{C}}, label=lst:mirath-sign, numbers=none]
int mirath_sign(..., uint8_t *sk) {
    ...
    mirath_matrix_decompress_secret_key(S, C, H, pk, (const uint8_t *)sk);
    ...
}
void mirath_matrix_decompress_secret_key(ff_t S[MIRATH_VAR_FF_S_BYTES], ff_t C[MIRATH_VAR_FF_C_BYTES], ..., const uint8_t *sk) {
    ...
    parse_secret_key(seed_sk, seed_pk, sk);
    mirath_matrix_expand_seed_secret_matrix(S, C, seed_sk);
    mirath_matrix_compute_y(y, S, C, H);
    ...
}
\end{lstlisting}

From Listing~\ref{lst:mirath-sign}, the function \texttt{parse\_secret\_key} ensures that all bytes allocated to \texttt{seed\_sk} are copied from the first sixteen secret bytes of \texttt{sk}.  
The function \texttt{mirath\_matrix\_expand\_seed\_secret\_matrix} then uses this secret data to construct the matrices \texttt{S} and \texttt{C}, which are subsequently passed to \texttt{mirath\_matrix\_compute\_y}.

\begin{lstlisting}[style=cstyle, caption={Demonstration of the cache-based side-channel leakage}, label=lst:mirath-ff-product-arith, numbers=none]
void mirath_matrix_compute_y(ff_t ...,  const ff_t S[MIRATH_VAR_FF_S_BYTES], const ff_t C[MIRATH_VAR_FF_C_BYTES], ...) {
    ...
    mirath_matrix_ff_product(..., S, C, ...);
    ...
}
void mirath_matrix_ff_product(..., const ff_t *matrix1, const ff_t *matrix2, ...) {
    mirath_matrix_ff_product_arith(..., matrix1, matrix2, ...);
}
static inline void mirath_matrix_ff_product_arith(ff_t *result, const ff_t *matrix1, const ff_t *matrix2, ...) {
    ...
    entry_i_k = mirath_matrix_ff_get_entry(matrix1, n_rows1, i, k);
    entry_k_j = mirath_matrix_ff_get_entry(matrix2, n_cols1, k, j);
    entry_i_j ^= mirath_ff_product(entry_i_k, entry_k_j);
    ...
}
static inline ff_t mirath_ff_product(const ff_t a, const ff_t b) {
    return mirath_ff_mult_table[a + 16 * b];
}
\end{lstlisting}

From Listing~\ref{lst:mirath-ff-product-arith}, the matrices are passed to the function \texttt{mirath\_matrix\_ff\_product}, which serves as a wrapper for \texttt{mirath\_matrix\_ff\_product\_arith}.  
In this context, the values \texttt{entry\_i\_k} and \texttt{entry\_k\_j}, computed from the matrix cells, are derived from secret data.

Within the function \texttt{mirath\_ff\_product}, these two components are used to compute an index that accesses a specific cell in the precomputed table \texttt{mirath\_ff\_mult\_table}.  
Because this index depends on secret values, the memory access becomes vulnerable to a cache-based side-channel attack.

\subsubsection{Non-Exploitable CT Violations} 
\label{sec:noexploit}

Conversely, while some alerts were manually confirmed to be true constant-time violations, they were deemed non-exploitable by the schemes' developers.
We detail examples of such alerts and the explanation for their non-exploitability below.

\paragraph{LESS} Besides the exploitable alerts in LESS round 2, we also found two non-exploitable alerts. LESS uses rejection sampling to sample integers within a certain range. If an integer is not in the considered range, there is a branch based on the value. Since the branch discards the value, there is no gain for an attacker, making the reported CT violation non-exploitable.

\paragraph{QR-UOV} Similarly, QR-UOV uses rejection sampling to generate the expanded secret seed, which triggers a CT violation. We consider this CT violation as non-exploitable. 

\paragraph{SNOVA}
After our initial report to the SNOVA authors, the round 2 code also flagged one control flow leak and three memory access leaks that are related to the lookup on \texttt{L\_J\_nibble} and \texttt{R\_tr\_J\_nibble}. These operations depend on \texttt{X\_in\_GF16Matrix}, which is randomly chosen and unique for each signing operation. As such, the developers believe that exploiting this property would be infeasible or, at the very least, significantly less efficient than existing known attacks.  

\paragraph{CROSS}
We found two control flow leaks and a memory access leak in the functions \texttt{csprng\_fp\_mat} and \texttt{csprng\_fz\_mat} of CROSS. They are used to sample matrices (\texttt{V} and \texttt{W}), which are actually public: they are reconstructed by the verifier by expanding the
seed in the public key. For this reason, an attacker would gain no
useful information by exploiting the control flow and memory access
variations, as they already possess the public key.  

\paragraph{MAYO}
The multiplication leaks flagged in MAYO are non-exploitable alerts, and the developers think they are unavoidable. We do not know if these leakages can be exploited as realistic vulnerabilities. 
However, we suggested that the developers replace the vulnerable multiplication with a logical \texttt{and} operation to avoid such leaks.  

\paragraph{PERK}
The developers reviewed the alert in PERK, confirming the function \texttt{sig\_perk\_perm\_set\_random} works as intended. While this function may take secret data as input (when called from the keygen or the parsing functions) and runs in variable time, this does not constitute an issue, as the function does not leak secret data. Indeed, one can see that whenever the function \texttt{sig\_perk\_perm\_gen\_given\_random\_input} triggers an early abort, the internal random buffer used is regenerated, hence the only information leaking is the knowledge that some failure happened (which is expected by design). Being able to relate the discarded random buffer to the newly generated one would constitute an attack against SHAKE.

\medskip 

\textbf{Conclusion.}
Our experiments demonstrate the toolchain’s effectiveness in helping identify critical time leakage security issues in PQDSS implementations.
The systematic analysis of constant-time violations on round 1 and round 2  NIST PQDSS scheme candidates revealed 5 exploitable vulnerabilities that were reported and fixed by the scheme's developers.
It also found 10 constant-time issues that were acknowledged (but not fixed) and will help design more robust future implementations of the schemes.
Finally, we uncovered 9 constant-time violations that were deemed not-exploitable, typically found in the implementation of rejection sampling.
This last point can help configure future analysis of similar schemes to focus on finding relevant alerts.
It is interesting to notice that not all exploitable vulnerabilities were raised by a single tool, a fact that demonstrates the importance of our toolchain.

\subsection{\textbf{RQ2}: Symbolic Tools against Statistical Tools}
\label{sec:rq2}
Symbolic tools are very good at localizing potential CT violations according to the CT policy. At the same time, this localization is the weakness of statistical tools, as they only return an evaluation of the whole algorithm, effectively as a single bit. The big advantage of statistical algorithms is that they do not require a CT policy, but simply evaluate the execution time, and therefore can emulate the path an attacker could take to attack the algorithm. In that sense, both tools work great in harmony. Ideally, the symbolic algorithm is used first to find CT violations, which are then fixed until no more issues can be found. After that, statistical algorithms should be used to validate that the CT policy accurately represents all potential timing leak sources in the system on which the code will be executed. A significant downside that can be seen in our evaluation regarding statistical tools is in the selection of the input classes that are compared against each other. 
Our approach was to adopt a generic strategy to determine whether a selected subset of bits from the private key correlates with the observed execution time.
However, the bits we have chosen in our evaluation were sampled without knowledge of the internals of the algorithm. 
This severely limits the expressiveness of the statistical test, as in our case, we can only say that certain key bits correlate or do not correlate with the execution time. It may be that other bits in the key do correlate, or that certain internal intermediate computation steps correlate with time. At the same time, testing more bits, or even all the bits, runs into the 'multiple-testing problem' \cite{Noble2009}, as each test has a chance to produce an incorrect result, doing more tests increases the odds of the result being incorrect. While compensating for this effect is possible with techniques like the Bonferroni correction~\cite{bonferroni1936}, the approaches of \dudect and \rtlf would result in a very conservative test decision. Traditional techniques like comparing a static private key with a random key to test all bits at the same time~\cite{tvla,tvlapublickey,hardwarepktvla} cannot be used against blackbox asymmetric algorithms, as they may have public key dependent execution time or use rejection sampling techniques that would create false positives in the evaluation.

\textbf{Conclusion.} Comparison of statistical and non-statistical tools for detecting constant-time violations in the NIST PQDSS implementations highlights their complementary strengths and limitations.
Symbolic tools, like \binsec, excel at pinpointing specific violations per the constant-time policy, while statistical tools, like \dudect and \rtlf, broadly assess execution time vulnerabilities without requiring a policy, though limited by generic input selection and multiple-testing issues. Their combined use—symbolic for localization, followed by statistical for validation—enhances comprehensive side-channel analysis, despite challenges in blackbox testing and false positives, informing robust evaluation strategies for post-quantum cryptography.

\subsection{\textbf{RQ3}: Quality of Constant-Time Tools Alerts}
\label{sec:rq3}

\newcommand{\nomul}{$^{-\texttt{MUL}}$} 

In this part, we are interested in comparing the capabilities of the symbolic tool \binsec and the dynamic tool \timecop to report alerts that match the policy they target, that is, no secret-dependent branches, memory accesses, or multiplications for \binsec and no secret-dependent branches, memory accesses for \timecop.

We also report the results of \binsec excluding multiplication alerts, which will be denoted by {\binsec~\nomul}.
We show in \autoref{tab:rq3alerts}, for each tool, the number of unique binary alerts and the number of unique related source code locations (obtained from the debug information of the digital signatures libraries).
We note that \timecop reports around 17 times more alerts than \binsec. 

These alerts can be mapped to 191 distinct source locations, around 7.3 times more than the alerts reported by \binsec (26 distinct source locations). Note that \timecop reported all alerts identified by \binsec~\nomul, and the additional true positive alerts \timecop raised, which \binsec missed, were never reached by \binsec due to scalability limitations.
This makes \timecop mostly useful on the subset of the code that was not reached by \binsec.

\begin{table}[]
    \centering
    \caption{Comparison of \binsec and \timecop Alerts.\label{tab:rq3alerts}}  

    Counts the number of unique alerts reported by each tool (unique binary context), the number of unique source locations that corresponds to at least one binary alert, and an estimate of the false positive rate of binary alerts (on a subset of 24 and 76 alerts reported by \binsec and \timecop resp.; across 14 instances covering 14 primitives). 

    \begin{tabular}{r|>{\centering\arraybackslash}m{2.5cm}|>{\centering\arraybackslash}m{2.5cm}|>{\centering\arraybackslash}m{2.5cm}|>{\centering\arraybackslash}m{2.5cm}}
                 & \textbf{\# Unique Binary Alerts} & \textbf{\# Unique Source Alerts} & \textbf{On Subset: False Positive Rate$^*$} & \textbf{On Subset: \# True Source Alerts$^*$} \\ \hline \hline
            \timecop   & 10247 & 191 &   67.1\% & 25 \\ \hline

            \binsec    &   598 &   25 &    0\%   & 25 \\ \hline
     \binsec \nomul    &   267 &   13 &    0\%   & 12 \\ \hline

    \end{tabular}
\end{table}

\medskip 

\textbf{Analysis.}  
As \binsec is sound (it has no false positives with respect to the constant-time policy it targets, excluding potential bugs in its implementation), we might expect that all its alerts exhibit real constant-time violations with respect to its configuration. 
The configuration we use, however, could in principle break the soundness of the analysis: when we replace AES instruction opcodes by a not fully equivalent replacement in the \binsec intermediate language, we might introduce paths that cannot exist in practice.

Similarly, as \timecop uses a taint mechanism, it will likely over-approximate actual constant-time violations (as additional program constraints may prevent dependencies).

For these reasons, we performed a manual check on 24 \binsec alerts and 76 \timecop alerts.
To obtain a representative estimate, we randomly chose alerts from analyses of one instance with small parameters for each primitive.
This represents 92.3\% of the \binsec alerts and 39.8\% of the \timecop alerts.
Among those, we found that 0\% of the \binsec alerts and $67.1\%$ of the \timecop alerts were false positives with respect to the corresponding constant-time policy (\emph{i.e.}, the outcome of the branching condition/memory access reported did not actually depend on the secret data).

This permits us to conclude that our chosen configuration of \binsec did not impact the soundness of its results on the PQDSS benchmark, and that it can provide a significantly more precise report than \timecop.

\medskip 

\textbf{Scalability.}
In total, \binsec reports 598 alerts on the benchmark while \timecop reports 10247 (17 times more).
This can be explained by the maximal depth reached by the analysis.
\binsec might require SMT queries to decide the feasibility of a path and whether a branch or memory access is secret-dependent or not.
As we are evaluating cryptography algorithms, the further we are in the execution, the less likely it is for the solver to be able to answer them in a reasonable time.
After this point, \binsec will miss any potential following vulnerability.
By contrast, as \timecop performs a dynamic execution on a single trace, it is able to explore the digital signature algorithm until the end, albeit only on a single trace.
On the subset where we performed he manual analysis, the last column of \autoref{tab:rq3alerts} shows that all the alerts that were raised by \binsec\nomul were also raised by \timecop, and all the remaining 13 alerts were in a location unreached by the \binsec analysis.

\medskip 

\textbf{Examples of false positives.}
Among the false positives we found in the \timecop alerts, all of them were overtaint cases, where while the branching condition or memory access uses a secret variable, its value is in practice sufficiently constrained to ensure that only a single outcome is possible.

Take for instance the code in \autoref{lst:timecop-cons-fp} that is flagged in two \timecop alerts (rbc\_53\_vec.c:109, rbc\_53\_vec\_set\_random\_full\_rank\_with\_one and rbc\_53\_elt.c:107, rbc\_53\_elt\_is\_zero) in \texttt{Ryde\_1f}. Our manual analysis first suggested it was legitimate, which developers confirmed.
Yet, further analysis from the \binsec result proved the opposite: bit-level constraints prevent constant-time violations. Notably, \texttt{acc == 0} evaluates to 0 regardless of \texttt{secret}. This pattern later repeats, which affects the earlier alert.

\begin{lstlisting}[style=cstyle, caption={Example of code leading to a \timecop false positive alert, inspired by a case from Ryde\_1f}, label=lst:timecop-cons-fp, firstnumber=5]
unsigned char secret;
int main(){
  int rank;
  unsigned char acc;
  acc = secret | 0x1;
  if (acc == 0)
    rank = 0;
  else
    rank = 1;
}
\end{lstlisting}

Take now the code of \autoref{lst:timecop-fp}, simplified to illustrate another alert (at ggm\_tree.c:176
in function ryde\_1f\_ggm\_tree\_get\_sibling\_path) from the \texttt{Ryde\_1f} instance. 
Here, we expect \texttt{v} to be secret-dependent. The bitwise and (\&) operation with the mask \texttt{0xff} and the addition make sure that the value of \texttt{x} is always between 4352 and  4607. As the loop will always iterate $\textrm{ceil}(\textrm{log}_2(x))$ times, and as this resolves to 13 for all numbers in the possible range, the number of iterations is invariant for any value of \texttt{v}. Although \binsec treats the loop as a constant-time implementation, \timecop raised an alert due to not taking these constraints into consideration. 

\begin{lstlisting}[style=cstyle, caption={Example of code leading to a \timecop false positive alert, inspired by a case from Ryde\_1f}, label=lst:timecop-fp, firstnumber=5]
volatile size_t v = 1;
#define N_1_MASK 0xff
#define TREE_LEAVES 4352
#define SEEDS_OFFSET (TREE_LEAVES - 1)
int main(void) {
    size_t x = v;
        x &= N_1_MASK;
        x += SEEDS_OFFSET;
    while (x > 0)
        x = (x-1)/2;
    return x;
}
\end{lstlisting}

\medskip 

\textbf{Conclusion.} \binsec thus provides a more precise evaluation of the constant-time policy evaluation than \timecop, which raises significantly more alerts, resulting in a high rate of false positive alerts.
However, the precision of \binsec has a cost in analysis difficulty that prevents it from evaluating later parts of the primitive implementation that \timecop is able to reach. 

In a practical setting, we thus advise using both tools, but to focus first on the alarms raised by \binsec as they are much more likely to be true CT violations, and then to focus on the alarms raised by  \timecop on the part of the code not explored by \binsec. 

\section{Discussion}
\textbf{Statistical Fault Localization.} One issue we faced when using statistical techniques is that we were unable to conveniently localize the source of the leak. Statistical techniques can only tell you that a leak is present, but cannot specify where the leak is coming from. The leak could be coming from the implementation, or worse, from a systematic bias in the measuring setup. This is a huge disadvantage compared to other tools, which can (oftentimes) pinpoint the leak to a specific line of code. Finding the root cause in such a diverse set of complex algorithms proves to be very challenging.

\textbf{Missing Declassification Information.} None of the analyzed algorithms clearly specifies when specific parts of the algorithm can be treated as non-confidential. For example, the signature algorithms generate a signature, which is, of course, secret-dependent, but the signature itself is, by definition, public. An algorithm that would change the encoding of an already computed signature in a non-CT way would result in a non-critical CT violation, as the signature before the conversion can already be considered public. Without analyzing each and every algorithm in detail, it is hard to judge for a new scheme whether a CT violation is actually meaningful. We, therefore, want to encourage authors to clearly state in their supporting documentation and code when the value is intentionally not treated as a secret anymore to allow for easier automated analysis.

\textbf{Invulnerability.} Many of the results of our manually investigated findings were not vulnerable, mostly due to the mentioned rejection sampling techniques. While this is hard to work with as an external auditor, as an author of the scheme, these false positives are easy to filter out. The tool then effectively provides a list of locations that require special attention from the implementer, which is valuable in ruling out implementation mistakes.

\textbf{Compilation of PQDSS primitives.}
To properly apply the selected tools to the implementations of the NIST PQDSS submissions, we had to create a pipeline that properly compiles and configures the candidates and tools according to the requirements of the selected tools.
This has proven to be tedious work and may, in some contexts, lead to the evaluation of a version of the scheme that is not the one the developers expect.
To improve the ease of setup of such analyses in the future, a standardized compilation entry point for the scheme would be of great help.

\section{Conclusion}
\label{sec:conclusions}

We presented our extensive efforts in analyzing a broad set of primitives submitted to NIST PQDSS Round 1 and Round 2 for constant-time programming compliance, using our in-house developed automated toolchain.  
In addition, we discussed how the findings from different tools were consolidated, manually verified, and communicated to the respective developers.  
We highlighted several timing leakages identified through this combined effort, which contributed to improving the side-channel resistance of the affected primitives.  
We also outlined how both developers and the standardization committee can benefit from our toolchain by rapidly assessing the timing leakage resistance of candidate schemes.  
Finally, we suggested directions for implementing primitives using more standardized APIs to enable more automated and scalable analysis.

\section*{Acknowledgments}
This work was partially supported by the ``France 2030'' government investment plan managed by the French National Research Agency, under the reference {ANR-22-PECY-0005}.



\bibliographystyle{ieeetr}
\bibliography{literature.bib}

\appendix

\subsection{\timecop Test Harness}

\begin{lstlisting}[style=cstyle, caption={Test harness for \timecop for a constant-time test of the \texttt{crypto\_sign()} function.}, label=lst:timecop]
#include <sys/types.h>
#include <unistd.h>
#include <string.h>
#include <stdlib.h>

#include "poison.h"
#include "api.h"
#include "toolchain_randombytes.h"

#define TIMECOP_NUMBER_OF_EXECUTION 1
#define max_message_length 3300

int main() {
	unsigned char *sm;
	unsigned long long smlen = 0;
	unsigned char *m;
	unsigned long long mlen = 0;
	unsigned char sk[CRYPTO_SECRETKEYBYTES] = {0};
	int result = 2 ; 
	for (int i = 0; i < TIMECOP_NUMBER_OF_EXECUTION; i++) {
		mlen = 33*(i+1);
		m = (unsigned char *)calloc(mlen, sizeof(unsigned char));
		sm = (unsigned char *)calloc(mlen+CRYPTO_BYTES, sizeof(unsigned char));
		ct_randombytes(m, mlen);
		unsigned char public_key[CRYPTO_PUBLICKEYBYTES] = {0};
		(void)crypto_sign_keypair(public_key, sk);

		poison(sk, CRYPTO_SECRETKEYBYTES * sizeof(unsigned char));
		result = crypto_sign(sm, &smlen, m, mlen, sk); 
		unpoison(sk, CRYPTO_SECRETKEYBYTES * sizeof(unsigned char));
		free(sm);
		free(m);
	}
	return result;
}
\end{lstlisting}

\end{document}

%% file: glossary.tex
\newglossaryentry{binsec}{
    name={Binsec},
    description={},
}

\newglossaryentry{flowtracker}{
    name={FlowTracker},
    description={},
}

\newglossaryentry{dudect}{
    name={Dudect},
    description={},
}

\newglossaryentry{ctgrind}{
    name={ctgrind},
    description={},
}

\newglossaryentry{timecop}{
    name={Timecop},
    description={},
}
\newglossaryentry{rtlf}{
    name={RTLF},
    description={},
}
\newcommand{\crash}{\scalebox{1.6}[1]{\textbf{\lightning}}}

\newcommand{\cmark}{\ding{51}}

%% file: comparison_table1.tex
\newcommand{\CTVrejsamp}{\faBomb}  
\newcommand{\CTVcare}{\faBomb}  
\newcommand{\CTVdontcare}{\faBomb} 
\newcommand{\truepositive}{\faBomb} 
\newcommand{\timeout}{\cmark} 
\newcommand{\reported}{R} 
\newcommand{\noalert}{\cmark}
\newcommand{\rawalert}{$\boldsymbol{!}$}
\newcommand{\notrelevant}{\faTimesCircle}
\newcommand{\notevaluated}{\textnormal{-}}
\newcommand{\nointegration}{\crash}

\begin{table*}

\begin{center}
        \tiny

\resizebox{\textwidth}{!}{
\begin{tabular}{ccc}
\begin{tabular}{r|r|cccc}
        \toprule
        & & B & D & R & T \\ \hline

        \multirow{21}{*}{\rotatebox{90}{Round 1}} 
        
& \texttt{SNOVA-24-5-16-4-esk} & \truepositive & \notevaluated & \notevaluated & \rawalert \\
& \texttt{SNOVA-24-5-16-4-ssk} & \rawalert & \notevaluated & \notevaluated & \rawalert \\
& \texttt{SNOVA-25-8-16-3-esk} & \rawalert & \notevaluated & \notevaluated & \rawalert \\
& \texttt{SNOVA-25-8-16-3-ssk} & \rawalert & \notevaluated & \notevaluated & \rawalert \\
& \texttt{SNOVA-28-17-16-2-esk} & \rawalert & \notevaluated & \notevaluated & \rawalert \\
& \texttt{SNOVA-28-17-16-2-ssk} & \rawalert & \notevaluated & \notevaluated & \rawalert \\
& \texttt{SNOVA-37-8-16-4-esk} & \rawalert & \notevaluated & \notevaluated & \rawalert \\
& \texttt{SNOVA-37-8-16-4-ssk} & \rawalert & \notevaluated & \notevaluated & \rawalert \\
& \texttt{SNOVA-43-25-16-2-esk} & \rawalert & \notevaluated & \notevaluated & \rawalert \\
& \texttt{SNOVA-43-25-16-2-ssk} & \rawalert & \notevaluated & \notevaluated & \rawalert \\
& \texttt{SNOVA-49-11-16-3-esk} & \rawalert & \notevaluated & \notevaluated & \rawalert \\
& \texttt{SNOVA-49-11-16-3-ssk} & \rawalert & \notevaluated & \notevaluated & \rawalert \\
& \texttt{SNOVA-60-10-16-4-esk} & \rawalert & \notevaluated & \notevaluated & \rawalert \\
& \texttt{SNOVA-60-10-16-4-ssk} & \rawalert & \notevaluated & \notevaluated & \rawalert \\
& \texttt{SNOVA-61-33-16-2-esk} & \rawalert & \notevaluated & \notevaluated & \rawalert \\
& \texttt{SNOVA-61-33-16-2-ssk} & \rawalert & \notevaluated & \notevaluated & \rawalert \\
& \texttt{SNOVA-66-15-16-3-esk} & \rawalert & \notevaluated & \notevaluated & \rawalert \\
& \texttt{SNOVA-66-15-16-3-ssk} & \rawalert & \notevaluated & \notevaluated & \rawalert \\
& \texttt{Preon128} & \truepositive & \notevaluated & \notevaluated & \truepositive \\
& \texttt{Preon192} & \rawalert & \notevaluated & \notevaluated & \rawalert \\
& \texttt{Preon256} & \rawalert & \notevaluated & \notevaluated & \rawalert \\
\texttt{--} & \texttt{------------------------------------} & \multicolumn{4}{c}{\texttt{------------------------------}} \\
\multirow{43}{*}{\rotatebox{90}{Round 2}} 
& \texttt{1\_RSDP\_BALANCED} & \CTVrejsamp & \noalert & \noalert & \notrelevant \\
& \texttt{1\_RSDPG\_BALANCED} & \CTVrejsamp & \noalert & \noalert & \notrelevant \\
& \texttt{1\_RSDPG\_SIG\_SIZE} & \CTVrejsamp & \noalert & \noalert & \notrelevant \\
& \texttt{1\_RSDPG\_SPEED} & \CTVrejsamp & \noalert & \noalert & \notrelevant \\
& \texttt{1\_RSDP\_SIG\_SIZE} & \timeout & \noalert & \noalert & \notrelevant \\
& \texttt{1\_RSDP\_SPEED} & \CTVrejsamp & \noalert & \noalert & \notrelevant \\
& \texttt{3\_RSDP\_BALANCED} & \CTVrejsamp & \noalert & \noalert & \notrelevant \\
& \texttt{3\_RSDPG\_BALANCED} & \CTVrejsamp & \noalert & \noalert & \notrelevant \\
& \texttt{3\_RSDPG\_SIG\_SIZE} & \CTVrejsamp & \noalert & \noalert & \notrelevant \\
& \texttt{3\_RSDPG\_SPEED} &  \CTVrejsamp & \noalert & \noalert & \notrelevant \\
& \texttt{3\_RSDP\_SIG\_SIZE} & \timeout & \noalert & \noalert & \notrelevant \\
& \texttt{3\_RSDP\_SPEED} & \timeout & \noalert & \noalert & \notrelevant \\
& \texttt{5\_RSDP\_BALANCED} & \timeout & \noalert & \noalert & \notrelevant \\
& \texttt{5\_RSDPG\_BALANCED} & \CTVrejsamp & \noalert & \noalert & \notrelevant \\
& \texttt{5\_RSDPG\_SIG\_SIZE} & \CTVrejsamp & \noalert & \noalert & \notrelevant \\
& \texttt{5\_RSDPG\_SPEED} & \CTVrejsamp & \noalert & \noalert & \notrelevant \\
& \texttt{5\_RSDP\_SIG\_SIZE} & \CTVrejsamp & \noalert & \noalert & \notrelevant \\
& \texttt{5\_RSDP\_SPEED} & \CTVrejsamp & \noalert & \noalert & \notrelevant  \\
& \texttt{less\_252\_192} & \CTVrejsamp & \noalert & \noalert & \truepositive \\
& \texttt{less\_252\_45} & \CTVrejsamp & \noalert & \noalert  & \truepositive \\
& \texttt{less\_252\_68} & \CTVrejsamp & \noalert & \rawalert  & \truepositive \\
& \texttt{less\_400\_102} & \CTVrejsamp & \noalert & \rawalert & \truepositive \\
& \texttt{less\_400\_220} & \CTVrejsamp & \noalert & \noalert & \truepositive \\
& \texttt{less\_548\_137} & \CTVrejsamp & \noalert & \noalert & \truepositive \\
& \texttt{less\_548\_345} & \CTVrejsamp & \noalert & \noalert & \truepositive \\
& \texttt{hawk256} & \reported & \noalert & \rawalert & \notrelevant \\
& \texttt{hawk512} & \reported & \noalert & \rawalert & \notrelevant \\
& \texttt{hawk1024} & \reported & \noalert & \rawalert & \notrelevant \\
& \texttt{mirath\_tcith\_1a\_fast} & \truepositive & \noalert & \noalert & \truepositive \\
& \texttt{mirath\_tcith\_1a\_short} & \truepositive & \noalert & \noalert & \truepositive \\
& \texttt{mirath\_tcith\_1b\_fast} & \truepositive & \noalert & \noalert & \rawalert \\
& \texttt{mirath\_tcith\_1b\_short} & \truepositive &  \noalert & \noalert & \rawalert \\
& \texttt{mirath\_tcith\_3a\_fast} & \truepositive &  \noalert & \noalert & \truepositive \\
& \texttt{mirath\_tcith\_3a\_short} & \truepositive & \noalert & \rawalert & \truepositive \\ 
& \texttt{mirath\_tcith\_3b\_fast} & \truepositive & \noalert & \rawalert & \rawalert \\
& \texttt{mirath\_tcith\_3b\_short} & \truepositive & \noalert & \rawalert & \rawalert \\
& \texttt{mirath\_tcith\_5a\_fast} & \truepositive & \noalert & \noalert & \truepositive \\ 
& \texttt{mirath\_tcith\_5a\_short} & \truepositive & \noalert & \noalert & \truepositive \\
& \texttt{mirath\_tcith\_5b\_fast} & \truepositive & \noalert & \noalert & \rawalert \\
& \texttt{mirath\_tcith\_5b\_short} & \truepositive & \noalert & \noalert & \rawalert \\
& \texttt{mqom2\_cat1\_gf256\_fast\_r3} & \timeout & \noalert & \noalert & \notrelevant \\
& \texttt{mqom2\_cat1\_gf256\_fast\_r5} & \timeout & \noalert & \noalert & \notrelevant \\
& \texttt{mqom2\_cat1\_gf256\_short\_r3} & \timeout & \noalert & \noalert & \notrelevant \\
&  &  & &  &  \\

        \bottomrule
\end{tabular}
&
\begin{tabular}{r|r|cccc}
        \toprule
& & B & D & R & T \\ \hline
\multirow{64}{*}{\rotatebox{90}{Round 2}} 

& \texttt{mqom2\_cat1\_gf256\_short\_r5} & \timeout & \noalert & \noalert & \notrelevant \\
& \texttt{mqom2\_cat1\_gf2\_fast\_r3} & \timeout & \noalert & \noalert & \notrelevant \\
& \texttt{mqom2\_cat1\_gf2\_fast\_r5} & \timeout & \noalert & \rawalert & \notrelevant \\
& \texttt{mqom2\_cat1\_gf2\_short\_r3} & \timeout & \noalert & \rawalert & \notrelevant \\
& \texttt{mqom2\_cat1\_gf2\_short\_r5} & \timeout & \noalert & \noalert & \notrelevant \\
& \texttt{mqom2\_cat3\_gf256\_fast\_r3} & \timeout & \noalert & \rawalert & \notrelevant \\
& \texttt{mqom2\_cat3\_gf256\_fast\_r5} & \timeout & \noalert & \rawalert & \notrelevant \\
& \texttt{mqom2\_cat3\_gf256\_short\_r3} & \timeout & \noalert & \noalert & \notrelevant \\
& \texttt{mqom2\_cat3\_gf256\_short\_r5} & \timeout & \noalert & \noalert & \notrelevant \\
& \texttt{mqom2\_cat3\_gf2\_fast\_r3} & \timeout & \noalert & \noalert & \notrelevant \\
& \texttt{mqom2\_cat3\_gf2\_fast\_r5} & \timeout & \noalert & \rawalert & \notrelevant \\
& \texttt{mqom2\_cat3\_gf2\_short\_r3} & \timeout & \noalert & \noalert & \notrelevant \\
& \texttt{mqom2\_cat3\_gf2\_short\_r5} & \timeout & \noalert & \noalert & \notrelevant \\
& \texttt{mqom2\_cat5\_gf256\_fast\_r3} & \timeout & \noalert & \noalert & \notrelevant \\
& \texttt{mqom2\_cat5\_gf256\_fast\_r5} & \timeout & \noalert & \noalert & \notrelevant \\
& \texttt{mqom2\_cat5\_gf256\_short\_r3} & \timeout & \noalert & \rawalert & \notrelevant \\
& \texttt{mqom2\_cat5\_gf256\_short\_r5} & \timeout & \noalert & \noalert & \notrelevant \\
& \texttt{mqom2\_cat5\_gf2\_fast\_r3} & \timeout & \noalert & \noalert & \notrelevant \\
& \texttt{mqom2\_cat5\_gf2\_fast\_r5} & \timeout & \noalert & \rawalert & \notrelevant \\
& \texttt{mqom2\_cat5\_gf2\_short\_r3} & \timeout & \noalert & \rawalert & \notrelevant \\
& \texttt{mqom2\_cat5\_gf2\_short\_r5} & \timeout & \noalert & \rawalert & \notrelevant \\
& \texttt{perk-192-short-5} & \timeout & \noalert & \noalert & \notrelevant \\
& \texttt{perk-192-fast-5} & \timeout & \noalert & \noalert & \notrelevant \\
& \texttt{perk-192-fast-3} & \timeout & \noalert & \noalert & \notrelevant \\
& \texttt{perk-192-short-3} & \timeout & \noalert & \noalert & \notrelevant \\
& \texttt{perk-128-fast-5} & \timeout & \noalert & \noalert & \notrelevant \\
& \texttt{perk-256-fast-3} & \timeout & \noalert & \noalert & \notrelevant \\
& \texttt{perk-128-fast-3} & \timeout & \noalert & \noalert & \notrelevant \\
& \texttt{perk-256-fast-5} & \timeout & \noalert & \noalert & \notrelevant \\
& \texttt{perk-128-short-3} & \timeout & \noalert & \noalert & \notrelevant \\
& \texttt{perk-128-short-5} & \timeout & \noalert & \rawalert & \notrelevant \\
& \texttt{perk-256-short-3} & \timeout & \noalert & \noalert & \notrelevant \\
& \texttt{perk-256-short-5} & \timeout & \noalert & \noalert & \notrelevant \\
& \texttt{ryde1f} & \CTVdontcare & \rawalert & \rawalert & \truepositive \\
& \texttt{ryde1s} & \CTVdontcare & \noalert & \rawalert & \truepositive \\
& \texttt{ryde3f} & \CTVdontcare & \noalert & \rawalert & \rawalert \\
& \texttt{ryde3s} & \CTVdontcare & \rawalert & \noalert & \rawalert \\
& \texttt{ryde5f} & \CTVdontcare & \noalert & \rawalert & \rawalert \\
& \texttt{ryde5s} & \CTVdontcare & \notevaluated & \notevaluated & \rawalert \\
& \texttt{SNOVA\_24\_5\_16\_4\_ESK} & \CTVdontcare & \notevaluated & \notevaluated & \rawalert \\
& \texttt{SNOVA\_24\_5\_16\_4\_SHAKE\_ESK} & \CTVdontcare & \notevaluated & \notevaluated & \rawalert \\
& \texttt{SNOVA\_24\_5\_16\_4\_SHAKE\_SSK} & \CTVdontcare & \notevaluated & \notevaluated & \rawalert \\
& \texttt{SNOVA\_24\_5\_16\_4\_SSK} & \CTVdontcare & \notevaluated & \notevaluated & \rawalert \\
& \texttt{SNOVA\_24\_5\_16\_5\_ESK} & \CTVdontcare & \notevaluated & \notevaluated & \rawalert \\
& \texttt{SNOVA\_24\_5\_16\_5\_SHAKE\_ESK} & \CTVdontcare & \notevaluated & \notevaluated & \rawalert \\
& \texttt{SNOVA\_24\_5\_16\_5\_SHAKE\_SSK} & \timeout & \notevaluated & \notevaluated & \rawalert \\
& \texttt{SNOVA\_24\_5\_16\_5\_SSK} & \CTVdontcare & \notevaluated & \notevaluated & \rawalert \\
& \texttt{SNOVA\_25\_8\_16\_3\_ESK} & \CTVdontcare & \notevaluated & \notevaluated & \rawalert \\
& \texttt{SNOVA\_25\_8\_16\_3\_SHAKE\_ESK} & \CTVdontcare & \notevaluated & \notevaluated & \rawalert \\
& \texttt{SNOVA\_25\_8\_16\_3\_SHAKE\_SSK} & \timeout & \notevaluated & \notevaluated & \rawalert \\
& \texttt{SNOVA\_25\_8\_16\_3\_SSK} & \timeout & \notevaluated & \notevaluated & \rawalert \\
& \texttt{SNOVA\_29\_6\_16\_5\_ESK} & \timeout & \notevaluated & \notevaluated & \rawalert \\
& \texttt{SNOVA\_29\_6\_16\_5\_SHAKE\_ESK} & \timeout & \notevaluated & \notevaluated  & \rawalert \\
& \texttt{SNOVA\_29\_6\_16\_5\_SHAKE\_SSK} & \timeout & \notevaluated & \notevaluated & \rawalert \\
& \texttt{SNOVA\_29\_6\_16\_5\_SSK} & \timeout & \notevaluated & \notevaluated & \rawalert \\
& \texttt{SNOVA\_37\_17\_16\_2\_ESK} & \CTVdontcare & \notevaluated & \notevaluated & \rawalert \\
& \texttt{SNOVA\_37\_17\_16\_2\_SHAKE\_ESK} & \CTVdontcare & \notevaluated & \notevaluated & \rawalert \\
& \texttt{SNOVA\_37\_17\_16\_2\_SHAKE\_SSK} & \timeout & \notevaluated & \notevaluated & \rawalert \\
& \texttt{SNOVA\_37\_17\_16\_2\_SSK} & \timeout & \notevaluated & \notevaluated & \rawalert \\
& \texttt{SNOVA\_37\_8\_16\_4\_ESK} & \CTVdontcare & \notevaluated & \notevaluated & \rawalert \\
& \texttt{SNOVA\_37\_8\_16\_4\_SHAKE\_ESK} & \CTVdontcare & \notevaluated & \notevaluated & \rawalert \\
& \texttt{SNOVA\_37\_8\_16\_4\_SHAKE\_SSK} & \timeout & \notevaluated & \notevaluated & \rawalert \\
& \texttt{SNOVA\_37\_8\_16\_4\_SSK} & \CTVdontcare & \notevaluated & \notevaluated & \rawalert \\
& \texttt{SNOVA\_49\_11\_16\_3\_ESK} & \CTVdontcare & \notevaluated & \notevaluated & \rawalert \\
& \texttt{SNOVA\_49\_11\_16\_3\_SHAKE\_ESK} & \CTVdontcare & \notevaluated & \notevaluated & \rawalert \\
& \texttt{SNOVA\_49\_11\_16\_3\_SHAKE\_SSK} & \timeout & \notevaluated & \notevaluated & \rawalert \\

        \bottomrule
    \end{tabular}
&
    \begin{tabular}{r|r|cccc}
        \toprule
& & B & D & R & T \\ \hline
\multirow{64}{*}{\rotatebox{90}{Round 2}} 

& \texttt{SNOVA\_49\_11\_16\_3\_SSK} & \timeout & \notevaluated & \notevaluated & \rawalert \\
& \texttt{SNOVA\_56\_25\_16\_2\_ESK} & \CTVdontcare & \notevaluated & \notevaluated & \rawalert \\
& \texttt{SNOVA\_56\_25\_16\_2\_SHAKE\_ESK} & \CTVdontcare & \notevaluated & \notevaluated & \rawalert \\
& \texttt{SNOVA\_56\_25\_16\_2\_SHAKE\_SSK} & \timeout & \notevaluated & \notevaluated & \rawalert \\
& \texttt{SNOVA\_56\_25\_16\_2\_SSK} & \timeout & \notevaluated & \notevaluated & \rawalert \\
& \texttt{SNOVA\_60\_10\_16\_4\_ESK} & \CTVdontcare & \notevaluated & \notevaluated & \rawalert \\
& \texttt{SNOVA\_60\_10\_16\_4\_SHAKE\_ESK} & \CTVdontcare & \notevaluated & \notevaluated & \rawalert \\
& \texttt{SNOVA\_60\_10\_16\_4\_SHAKE\_SSK} & \timeout & \notevaluated & \notevaluated & \rawalert \\
& \texttt{SNOVA\_60\_10\_16\_4\_SSK} & \CTVdontcare & \notevaluated & \notevaluated & \rawalert \\
& \texttt{SNOVA\_66\_15\_16\_3\_ESK} & \CTVdontcare & \notevaluated & \notevaluated & \rawalert \\
& \texttt{SNOVA\_66\_15\_16\_3\_SHAKE\_ESK} & \CTVdontcare & \notevaluated & \notevaluated & \rawalert \\
& \texttt{SNOVA\_66\_15\_16\_3\_SHAKE\_SSK} & \timeout & \notevaluated & \notevaluated & \rawalert \\
& \texttt{SNOVA\_66\_15\_16\_3\_SSK} & \timeout & \notevaluated & \notevaluated & \rawalert \\
& \texttt{SNOVA\_75\_33\_16\_2\_ESK} & \CTVdontcare & \notevaluated & \notevaluated & \rawalert \\
& \texttt{SNOVA\_75\_33\_16\_2\_SHAKE\_ESK} & \CTVdontcare & \notevaluated & \notevaluated & \rawalert \\
& \texttt{SNOVA\_75\_33\_16\_2\_SHAKE\_SSK} & \timeout & \notevaluated & \notevaluated & \rawalert \\
& \texttt{SNOVA\_75\_33\_16\_2\_SSK} & \timeout & \notevaluated & \notevaluated & \rawalert \\
& \texttt{sqisign\_lvl3} & \timeout & \noalert & \rawalert & \rawalert \\
& \texttt{sqisign\_lvl5} & \timeout & \noalert & \rawalert & \rawalert \\
& \texttt{sqisign\_lvl1} & \timeout & \noalert & \rawalert & \rawalert \\
& \texttt{mayo\_1} & \CTVdontcare & \noalert & \noalert & \notrelevant \\
& \texttt{mayo\_2} & \CTVdontcare & \noalert & \noalert & \notrelevant \\
& \texttt{mayo\_3} & \CTVdontcare & \noalert & \noalert & \notrelevant \\
& \texttt{mayo\_5} & \CTVdontcare & \noalert & \noalert & \notrelevant \\
& \texttt{sdith\_cat1\_fast} & \timeout & \noalert & \noalert & \rawalert \\
& \texttt{sdith\_cat1\_short} & \timeout & \noalert & \noalert & \rawalert \\
& \texttt{sdith\_cat3\_fast} & \timeout & \noalert & \noalert & \rawalert \\
& \texttt{sdith\_cat3\_short} & \timeout & \noalert & \noalert & \rawalert \\
& \texttt{sdith\_cat5\_fast} & \timeout & \noalert & \noalert & \rawalert \\
& \texttt{sdith\_cat5\_short} & \timeout & \noalert & \noalert & \rawalert \\
& \texttt{qruov1q127L3v156m54} & \CTVrejsamp & \noalert & \noalert & \rawalert \\
& \texttt{qruov1q31L10v600m70} & \CTVrejsamp & \noalert & \noalert & \rawalert \\
& \texttt{qruov1q31L3v165m60} & \CTVrejsamp & \noalert & \noalert & \rawalert \\
& \texttt{qruov1q7L10v740m100} & \CTVrejsamp & \noalert & \noalert & \rawalert \\
& \texttt{qruov3q127L3v228m78} & \CTVrejsamp & \noalert & \noalert & \rawalert \\
& \texttt{qruov3q31L10v890m100} & \CTVrejsamp & \noalert & \noalert & \rawalert \\
& \texttt{qruov3q31L3v246m87} & \CTVrejsamp & \noalert & \noalert & \rawalert \\
& \texttt{qruov3q7L10v1100m140} & \CTVrejsamp & \noalert & \noalert & \rawalert \\
& \texttt{qruov5q127L3v306m105} & \CTVrejsamp & \noalert & \noalert & \rawalert \\
& \texttt{qruov5q31L10v1120m120} & \CTVrejsamp & \noalert & \noalert & \rawalert \\
& \texttt{qruov5q31L3v324m114} & \CTVrejsamp & \noalert & \noalert & \rawalert \\
& \texttt{qruov5q7L10v1490m190} & \CTVrejsamp & \notevaluated & \noalert & \rawalert \\
& \texttt{uov\_OV16\_160\_64\_4rAES\_classic} & \timeout & \notevaluated & \notevaluated & \rawalert \\
& \texttt{uov\_OV16\_160\_64\_4rAES\_pkc} & \timeout & \notevaluated & \notevaluated & \rawalert \\
& \texttt{uov\_OV16\_160\_64\_4rAES\_pkc+skc} & \timeout & \notevaluated & \notevaluated & \rawalert \\
& \texttt{uov\_OV256\_112\_44\_4rAES\_classic} & \timeout & \notevaluated & \notevaluated & \rawalert \\
& \texttt{uov\_OV256\_112\_44\_4rAES\_pkc} & \timeout & \notevaluated & \notevaluated & \rawalert \\
& \texttt{uov\_OV256\_112\_44\_4rAES\_pkc+skc} & \timeout & \notevaluated & \notevaluated & \rawalert \\
& \texttt{uov\_OV256\_184\_72\_4rAES\_classic} & \timeout & \notevaluated & \notevaluated & \rawalert \\
& \texttt{uov\_OV256\_184\_72\_4rAES\_pkc} & \timeout & \notevaluated & \notevaluated & \rawalert \\
& \texttt{uov\_OV256\_184\_72\_4rAES\_pkc+skc} & \timeout & \notevaluated & \notevaluated & \rawalert \\
& \texttt{uov\_OV256\_244\_96\_4rAES\_classic} & \timeout & \notevaluated & \notevaluated & \rawalert \\
& \texttt{uov\_OV256\_244\_96\_4rAES\_pkc} & \timeout & \notevaluated & \notevaluated & \rawalert \\
& \texttt{uov\_OV256\_244\_96\_4rAES\_pkc+skc} & \timeout & \notevaluated & \notevaluated & \rawalert \\
& \texttt{faest\_128f} & \nointegration & \nointegration & \nointegration & \nointegration \\
& \texttt{faest\_128s} & \nointegration & \nointegration & \nointegration & \nointegration \\
& \texttt{faest\_192f} & \nointegration & \nointegration & \nointegration & \nointegration \\
& \texttt{faest\_192s} & \nointegration & \nointegration & \nointegration & \nointegration \\
& \texttt{faest\_256f} & \nointegration & \nointegration & \nointegration & \nointegration \\
& \texttt{faest\_256s} & \nointegration & \nointegration & \nointegration & \nointegration \\
& \texttt{faest\_em\_128f} & \nointegration & \nointegration & \nointegration & \nointegration \\
& \texttt{faest\_em\_128s} & \nointegration & \nointegration & \nointegration & \nointegration \\
& \texttt{faest\_em\_192f} & \nointegration & \nointegration & \nointegration & \nointegration \\
& \texttt{faest\_em\_192s} & \nointegration & \nointegration & \nointegration & \nointegration \\
& \texttt{faest\_em\_256f} & \nointegration & \nointegration & \nointegration & \nointegration \\
& \texttt{faest\_em\_256s} & \nointegration & \nointegration & \nointegration & \nointegration \\

        \bottomrule
    \end{tabular}
\\
\end{tabular}
}
\vspace{1em}
\caption{Overview of our executed experiments. A \noalert~indicates that the experiment was run successfully with no reported issues. A \rawalert~indicates that the respective tool reported an issue in the implementation, which we did not manually analyze. A \truepositive~symbol indicates that the tool reported a vulnerability, and manual analysis confirmed the vulnerability as a true positive. A \notrelevant~indicates that the tool reported a vulnerability, but manual analysis found the reported issue to be a false positive. Last, a \nointegration~indicates an issue with integrating the candidate into the toolchain, meaning no analysis could be performed. A \notevaluated~means we have not performed this experiment yet. A \reported~symbol indicates that the tool has reported issues to the developers, but acknowledgment is still pending. Multiple symbols in a cell refer to different discovered CT violations. \textbf{B} is \binsec, \textbf{D} is \dudect, \textbf{R} is \rtlf, \textbf{T} is \timecop.}
    \label{tab:results}
\end{center}    
\end{table*}

%% file: comparison_table2_nist.tex
\begin{table*}[ht]
\scriptsize
\begin{center}
\begin{tabular}{lcccccccccccccc}
\toprule
\multirow{3}{*}{Candidate} &
    \multicolumn{4}{c}{Total Alerts} &
    \multicolumn{6}{c}{Flagged Location} &
    \multicolumn{4}{c}{Disclosure Status} \\

\cmidrule(lr){2-5} \cmidrule(lr){6-11} \cmidrule(lr){12-15}

    & \multicolumn{2}{c}{Symbolic Tools} & \multicolumn{2}{c}{Statical Tools}
    & \multicolumn{3}{c}{Source Locations} & \multicolumn{3}{c}{Reported Locations}
    & \multicolumn{2}{c}{Acknowledged} & \multicolumn{2}{c}{Not Acknowledged} \\

\cmidrule(lr){2-3} \cmidrule(lr){4-5} \cmidrule(lr){6-8} \cmidrule(lr){9-11} \cmidrule(lr){12-13} \cmidrule(lr){14-15}

    & \textbf{B} & \textbf{T} & \textbf{D} & \textbf{R}
    & \textbf{B} & \textbf{T} & Total & \textbf{B} & \textbf{T} & Total
    & Fixed & Remain & No answer & Rejected \\

\midrule
SNOVA (1) & 1 & 160 & - & - &   1  &  9  &  10  &   1  &  0   &   1  &  1   &   0  &   0  &   0  \\
Preon & 1 & 100 & - & - &   1  &  11  &  11  &   1  &  0   &   1  &  1   &   0  &   0  &   0  \\
CROSS   & 17 & 167 & 9 & 17 &  2   &  15   &  15   &  2   &  2   &  2   &   0  &  2   &   0  &  0   \\ 
LESS    &  7 &  1068 & 0 &  1 &   1  &  41   &   41  &  1   &  4   &   4  &  2   &   0  &   0  &  2   \\ 
HAWK    &  6 &  96 & 0 &  0 &   2  &  13   &   13  &  2   &  1   &  2   &   0  &   0  &   2  &   0  \\ 
Mirath  &  10 & 130 & 0 &  2 &  1   &  10    &  10 &  1  &  1   &   1  &   1  &   0  &  0   &  0  \\
MQOM    & 0  & 240 & 0 &  0 &   0  &  3   &  3   &  0   &  0   &  0   &  0   &  0   &  0   &  0   \\
PERK    & 0 & 1835 & 0 &  1 &   0  &  22   &   22  &  0   &  1   &   1  &  0   &  1   &  0   &   0  \\
RYDE    &  18 &  54 & 0 &  1 &  4   &   10  &   13  &  1   &   6  &  6   &  0   &   0  &   0  &  6   \\
SNOVA (2)  & 447 & 210 & 8 & 31 &   7  &  26   &   29   &   1  &  1   &   1  &  0   &   1  &   0  &   0  \\
SQIsign & 0  &  1118 & 0 &  0 &  0  &  6   &  6   &   0  &  1   &  1   &  0   &  0   &  0   &   1  \\
MAYO    & 80  &  4 & 0 &  0 &  5   &  1   &  6   &   5  &  0   & 5    &  0   &   5  &  0   &    0 \\
SDitH   &  0 &  461 & 0 &  0 &   0  &  17   &   17  &   0  &  0   &  0   &   0  &  0   &   0  &   0  \\
QR-UOV   & 11 & 246 & 0 &  8 &   1  &  6   &  6   &  1    & 1  &  1  &  0   &  1  &  0   & 0  \\ 
UOV   &  0 & 12 & 2 &  3 &   0  &  1   &  1   & 0   &  0   &  0   &    0 &  0   &  0   &   0  \\
FAEST   &  - &  - & - &  - &   - &   - &   - &   - &   - &   - &   - &   - &   - &   - \\
\bottomrule
\end{tabular}

\vspace{2mm}
\caption{Analysis of Findings by Tool and Candidate}
\label{tab:report}
\end{center}
\end{table*}


%% file: main.bbl
\begin{thebibliography}{10}

\bibitem{nist_pqc}
{National Institute of Standards and Technology (NIST)}, ``Post-quantum
  cryptography standardization.'' NIST Computer Security Resource Center, 2024.
\newblock Accessed: 2025-4-28.

\bibitem{FIPS-203}
N.~I. of~Standards and T.~(NIST), ``Fips 203: Module-lattice-based
  key-encapsulation mechanism standard.'' U.S. Department of Commerce, National
  Institute of Standards and Technology, 2024.

\bibitem{nist-pqc-selected}
{National Institute of Standards and Technology (NIST)}, ``Post-quantum
  cryptography selected algorithms.'' NIST Computer Security Resource Center,
  2024.
\newblock Accessed: 2025-4-28.

\bibitem{FIPS-204}
N.~I. of~Standards and T.~(NIST), ``Fips 204: Module-lattice-based digital
  signature standard.'' U.S. Department of Commerce, National Institute of
  Standards and Technology, 2024.

\bibitem{FIPS-205}
N.~I. of~Standards and T.~(NIST), ``Fips 205: Stateless hash-based digital
  signature standard.'' U.S. Department of Commerce, National Institute of
  Standards and Technology, 2024.

\bibitem{pqc-dig-sig}
{National Institute of Standards and Technology (NIST)}, ``Post-quantum
  cryptography: Additional digital signature schemes.'' NIST Computer Security
  Resource Center, 2023.
\newblock Accessed: 2025-4-28.

\bibitem{NIST-IR-8528}
G.~Alagic, M.~Bros, P.~Ciadoux, D.~Cooper, Q.~Dang, T.~Dang, J.~Kelsey,
  J.~Lichtinger, Y.-K. Liu, C.~Miller, D.~Moody, R.~Peralta, R.~Perlner,
  A.~Robinson, H.~Silberg, D.~Smith-Tone, and N.~Waller, ``Status report on the
  first round of the additional digital signature schemes for the nist
  post-quantum cryptography standardization process.'' U.S. Department of
  Commerce, National Institute of Standards and Technology, 2024.

\bibitem{nist-pqc-dig-sig-cfp-2022}
{National Institute of Standards and Technology}, ``Call for additional digital
  signature schemes for the post-quantum cryptography standardization
  process.''
  \url{https://csrc.nist.gov/csrc/media/Projects/pqc-dig-sig/documents/call-for-proposals-dig-sig-sept-2022.pdf},
  Sept. 2022.
\newblock Accessed: 2025-04-18.

\bibitem{nist-pqc-security-criteria}
{National Institute of Standards and Technology}, ``Post-quantum cryptography
  -- security (evaluation criteria).''
  \url{https://csrc.nist.gov/projects/post-quantum-cryptography/post-quantum-cryptography-standardization/evaluation-criteria/security-(evaluation-criteria)},
  2024.
\newblock Accessed: 2025-04-18.

\bibitem{timingntruencrypt2007}
J.~H. Silverman and W.~Whyte, ``Timing attacks on ntruencrypt via variation in
  the number of hash calls,'' in {\em Proceedings of the 7th Cryptographers'
  Track at the RSA Conference on Topics in Cryptology}, CT-RSA'07, (Berlin,
  Heidelberg), p.~208–224, Springer-Verlag, 2007.

\bibitem{cryptoeprint:2016/300}
L.~G. Bruinderink, A.~Hülsing, T.~Lange, and Y.~Yarom, ``Flush, gauss, and
  reload -- a cache attack on the bliss lattice-based signature scheme.''
  Cryptology ePrint Archive, Paper 2016/300, 2016.
\newblock \url{https://eprint.iacr.org/2016/300}.

\bibitem{cryptoeprint:2017/505}
T.~Espitau, P.-A. Fouque, B.~Gerard, and M.~Tibouchi, ``Side-channel attacks on
  bliss lattice-based signatures -- exploiting branch tracing against
  strongswan and electromagnetic emanations in microcontrollers.'' Cryptology
  ePrint Archive, Paper 2017/505, 2017.
\newblock \url{https://eprint.iacr.org/2017/505}.

\bibitem{kyberslash}
``Kyberslash: division timings depending on secrets in kyber software.''
  \url{https://kyberslash.cr.yp.to/}.
\newblock Accessed: 25.01.2024.

\bibitem{dudect16}
O.~Reparaz, J.~Balasch, and I.~Verbauwhede, ``Dude, is my code constant
  time?.'' Cryptology ePrint Archive, Paper 2016/1123, 2016.

\bibitem{mona12}
S.~Schinzel, {\em Unintentional and Hidden Information Leaks in Networked
  Software Applications}.
\newblock doctoralthesis, Friedrich-Alexander-Universit{\"a}t
  Erlangen-N{\"u}rnberg (FAU), 2012.

\bibitem{tlsfuzzer23}
H.~Kario, ``Everlasting robot: the marvin attack.'' Computer Security --
  ESORICS 2023, 2023.

\bibitem{rtlf}
M.~Dunsche, M.~Maehren, N.~Erinola, R.~Merget, N.~Bissantz, J.~Somorovsky, and
  J.~Schwenk, ``With great power come great side channels: Statistical timing
  {Side-Channel} analyses with bounded type-1 errors,'' in {\em 33rd USENIX
  Security Symposium (USENIX Security 24)}, (Philadelphia, PA), pp.~6687--6704,
  USENIX Association, Aug. 2024.

\bibitem{blazer}
T.~Antonopoulos, P.~Gazzillo, M.~Hicks, E.~Koskinen, T.~Terauchi, and S.~Wei,
  ``Decomposition instead of self-composition for proving the absence of timing
  channels,'' in {\em Proceedings of the 38th ACM SIGPLAN Conference on
  Programming Language Design and Implementation}, PLDI 2017, (New York, NY,
  USA), p.~362–375, Association for Computing Machinery, 2017.

\bibitem{cacheaudit13}
G.~Doychev, D.~Feld, B.~Köpf, L.~Mauborgne, and J.~Reineke, ``{CacheAudit}: A
  tool for the static analysis of cache side channels.'' Cryptology ePrint
  Archive, Paper 2013/253, 2013.

\bibitem{flowtracker}
B.~Rodrigues, F.~M.~Q. Pereira, and D.~F. Aranha, ``Sparse representation of
  implicit flows with applications to side-channel detection,'' in {\em
  Proceedings of the 25th International Conference on Compiler Construction,
  {CC} 2016, Barcelona, Spain, March 12-18, 2016} (A.~Zaks and M.~V.
  Hermenegildo, eds.), pp.~110--120, {ACM}, 2016.

\bibitem{ctllvm}
Z.~Zhang and G.~Barthe, ``{CT}-{LLVM}: Automatic large-scale constant-time
  analysis.'' Cryptology {ePrint} Archive, Paper 2025/338, 2025.

\bibitem{ctgrind10}
A.~Langley, ``ctgrind: Checking that functions are constant time with
  valgrind.''

\bibitem{microwalk18}
J.~Wichelmann, A.~Moghimi, T.~Eisenbarth, and B.~Sunar, ``Microwalk: {A}
  framework for finding side channels in binaries,'' in {\em {ACSAC}},
  pp.~161--173, {ACM}, 2018.

\bibitem{data18}
S.~Weiser, A.~Zankl, R.~Spreitzer, K.~Miller, S.~Mangard, and G.~Sigl,
  ``{DATA}--differential address trace analysis: Finding address-based
  side-channels in binaries,'' in {\em Proceedings of the 27th USENIX
  Conference on Security Symposium}, SEC'18, (USA), p.~603–620, USENIX
  Association, 2018.

\bibitem{timecop}
M.~Neikes, ``Timecop: Automated dynamic analysis for timing side-channels.''

\bibitem{binsecrel19}
L.~Daniel, S.~Bardin, and T.~Rezk, ``Binsec/rel: Efficient relational symbolic
  execution for constant-time at binary-level,'' in {\em {SP}}, pp.~1021--1038,
  {IEEE}, 2020.

\bibitem{cached17}
S.~Wang, P.~Wang, X.~Liu, D.~Zhang, and D.~Wu, ``{CacheD}: Identifying
  cache-based timing channels in production software,'' in {\em Proceedings of
  the 26th USENIX Conference on Security Symposium}, SEC'17, (USA),
  p.~235–252, USENIX Association, 2017.

\bibitem{notthathardtomitigate21}
J.~Jancar, M.~Fourné, D.~D.~A. Braga, M.~Sabt, P.~Schwabe, G.~Barthe, P.-A.
  Fouque, and Y.~Acar, ``“they’re not that hard to mitigate”: What
  cryptographic library developers think about timing attacks.'' Cryptology
  ePrint Archive, Paper 2021/1650, 2021.

\bibitem{GeimerVRDBM23}
A.~Geimer, M.~Vergnolle, F.~Recoules, L.~Daniel, S.~Bardin, and C.~Maurice, ``A
  systematic evaluation of automated tools for side-channel vulnerabilities
  detection in cryptographic libraries,'' in {\em {CCS}}, pp.~1690--1704,
  {ACM}, 2023.

\bibitem{timingRsaDhKocher96}
P.~C. Kocher, ``Timing attacks on implementations of diffie-hellman, {RSA},
  {DSS}, and other systems,'' in {\em Advances in Cryptology - {CRYPTO} '96,
  16th Annual International Cryptology Conference, Santa Barbara, California,
  USA, August 18-22, 1996, Proceedings} (N.~Koblitz, ed.), vol.~1109 of {\em
  Lecture Notes in Computer Science}, pp.~104--113, Springer, 1996.

\bibitem{DBLP:conf/ccs/ShivakumarBGLP22}
B.~A. Shivakumar, G.~Barthe, B.~Gr{\'{e}}goire, V.~Laporte, and S.~Priya,
  ``Enforcing fine-grained constant-time policies,'' in {\em {CCS}},
  pp.~83--96, {ACM}, 2022.

\bibitem{DBLP:conf/uss/DanielBNBRP23}
L.~Daniel, M.~Bognar, J.~Noorman, S.~Bardin, T.~Rezk, and F.~Piessens,
  ``Prospect: Provably secure speculation for the constant-time policy,'' in
  {\em {USENIX} Security Symposium}, pp.~7161--7178, {USENIX} Association,
  2023.

\bibitem{groot15}
W.~de~Groot, ``A performance study of {X25519} on {Cortex-M3} and {M4}.''
  Master thesis, 9 2015.

\bibitem{hertzbleed22}
Y.~Wang, R.~Paccagnella, E.~He, H.~Shacham, C.~W. Fletcher, and D.~Kohlbrenner,
  ``Hertzbleed: Turning power side-channel attacks into remote timing attacks
  on x86,'' in {\em Proceedings of the USENIX Security Symposium (USENIX)},
  2022.

\bibitem{LW-toolchain}
A.~B. Hansen, E.~H. Nielsen, and M.~Eskildsen, ``Toolchain for timing leakage
  analysis of nist lightweight crypto candidates,'' 2020.

\bibitem{supercop}
T.~L. Daniel J.~Bernstein, ``e{BACS}: {ECRYPT} benchmarking of cryptographic
  systems.''

\bibitem{cryptoeprint:2024/1049}
D.~J. Bernstein, K.~Bhargavan, S.~Bhasin, A.~Chattopadhyay, T.~K. Chia, M.~J.
  Kannwischer, F.~Kiefer, T.~Paiva, P.~Ravi, and G.~Tamvada, ``{KyberSlash}:
  Exploiting secret-dependent division timings in kyber implementations.''
  Cryptology {ePrint} Archive, Paper 2024/1049, 2024.

\bibitem{supercop-sign-benchmarks}
D.~J. Bernstein, ``Supercop: List of public-key signature systems measured.''
  \url{https://bench.cr.yp.to/primitives-sign.html}, 2025.
\newblock Accessed: 2025-05-31.

\bibitem{nist-pqc-forum-2024}
B.~LaMacchia, ``Benchmark planning for on-ramp?.''
  \url{https://groups.google.com/a/list.nist.gov/g/pqc-forum/c/hRTPcxHohPM/m/dUhQEUNvCAAJ},
  April 2024.
\newblock Posted on the NIST PQC Forum.

\bibitem{Langley2010checkvalgrind}
A.~Langley, ``{Checking that functions are constant time with Valgrind}.''
  \url{https://www.imperialviolet.org/2010/04/01/ctgrind.html}, April 2010.
\newblock Accessed: 29-05-2025.

\bibitem{DanielBR23}
L.~Daniel, S.~Bardin, and T.~Rezk, ``Binsec/rel: Symbolic binary analyzer for
  security with applications to constant-time and secret-erasure,'' {\em {ACM}
  Trans. Priv. Secur.}, vol.~26, no.~2, pp.~11:1--11:42, 2023.

\bibitem{Sachs.1984}
L.~Sachs, {\em Applied Statistics: A Handbook of Techniques}.
\newblock Springer Series in Statistics, New York, NY: {Springer New York},
  2~ed., 1984.

\bibitem{Noble2009}
W.~S. Noble, ``How does multiple testing correction work?,'' {\em Nature
  Biotechnology}, vol.~27, no.~12, pp.~1135--1137, 2009.

\bibitem{bonferroni1936}
C.~Bonferroni, ``Teoria statistica delle classi e calcolo delle probabilita,''
  {\em Pubblicazioni del R istituto superiore di scienze economiche e
  commericiali di firenze}, vol.~8, pp.~3--62, 1936.

\bibitem{tvla}
T.~Lev-Ami and M.~Sagiv, ``Tvla: A system for implementing static analyses,''
  in {\em Static Analysis} (J.~Palsberg, ed.), (Berlin, Heidelberg),
  pp.~280--301, Springer Berlin Heidelberg, 2000.

\bibitem{tvlapublickey}
M.~Tunstall and G.~Goodwill, ``Applying tvla to public key cryptographic
  algorithms,'' {\em IACR Cryptol. ePrint Arch.}, vol.~2016, p.~513, 2016.

\bibitem{hardwarepktvla}
A.~Jayasena, E.~Andrews, and P.~Mishra, ``Tvla*: Test vector leakage assessment
  on hardware implementations of asymmetric cryptography algorithms,'' {\em
  IEEE Transactions on Very Large Scale Integration (VLSI) Systems}, vol.~31,
  no.~9, pp.~1269--1279, 2023.

\end{thebibliography}
